%% file: CanonicalTensorProducts.tex
\documentclass[aip,jcp,numerical,longbibliography,reprint]{revtex4-1}
\usepackage{color}
\usepackage{graphicx}
\usepackage{threeparttable}
\usepackage{longtable}
\usepackage{enumerate}
\usepackage{amsfonts,amssymb}
\usepackage{amsmath}
\usepackage{stmaryrd}
\usepackage{dcolumn}
\usepackage{bm}
\usepackage{bbm}
\usepackage{algpseudocode}
\usepackage{algorithm}
\usepackage[all,cmtip]{xy}
\usepackage{url}
\usepackage{multirow}
\usepackage{array}

\bibpunct{}{}{,}{s}{}{\textsuperscript{,}}  

\input{conventions.tex}

\begin{document}

\title{
Classifications and canonical forms of tensor product expressions in the presence of permutation symmetries
}
\author{Zhendong Li,$^1$
\footnote{To whom correspondence should be addressed: zhendongli2008@gmail.com}
\footnote{Present address: Department of Chemistry, Princeton University, Princeton, New Jersey 08544, USA}
Sihong Shao,$^2$
\footnote{To whom correspondence should be addressed: sihong@math.pku.edu.cn}
and Wenjian Liu$^1$
\footnote{To whom correspondence should be addressed: liuwj@pku.edu.cn}}
\bigskip
\affiliation{\footnotesize{$^1$ Beijing National Laboratory for
Molecular Sciences, Institute of Theoretical and Computational
Chemistry, State Key Laboratory of Rare Earth Materials Chemistry
and Applications, College of Chemistry and Molecular Engineering,
and Center for Computational Science and Engineering, Peking
University, Beijing 100871, People's Republic of China\\
$^2$ LMAM and School of Mathematical Sciences, Peking University,
Beijing 100871, People's Republic of China.}}
\begin{abstract}
Complicated mathematical equations involving products of tensors with permutation symmetries,
frequently encountered in fields such as general relativity and quantum chemistry (e.g., equations in high-order coupled cluster theories), require computer-based automatic derivations and manipulations.
In these processes, a key step is the collection of tensor product terms
that can be found identical by utilizing permutation symmetries of tensors or
relabeling dummy indices, which is usually achieved by defining a \emph{canonical
form} for tensor product expressions. However, the problem of finding a canonical form is
nontrivial, and can be potentially of \emph{exponential} cost in the number of indices.
In this work, we provided a general solution to this tensor canonicalization problem.
First, we developed a complete group theoretical classification of all possibly generated tensor products,
from which an unambiguous definition of \emph{canonical form} can be derived
for arbitrary tensor products. Second, to make such theoretical definition of practical use, inspired by
diagrammatic methods in perturbation theory as well as tensor network diagrams,
we proposed a graphical reformulation of the canonicalization problem,
which leads to an efficient algorithm to compute the canonical form based on the graph representation of tensor products. The algorithm combines the classical backtrack search for permutation groups and the concept of partitions used in graph isomorphism algorithms for more efficient pruning, which renders the size of the search tree
scale \emph{polynomially} rather than exponentially in difficult cases for existing algorithms.
It allows to compute both the canonical form of a tensor product and its automorphism group.
Besides, for a tensor product with external indices, its permutation symmetry group can be determined
using the same algorithm through a homomorphism, that is, as the quotient group of the automorphism group for the
corresponding \emph{externally and internally unlabeled graph (skeleton)}
with respect to that for the \emph{externally labeled and internally
unlabeled graph (diagram)}.
\end{abstract}
\maketitle

\tableofcontents

\section{Introduction}
In the most broad sense, tensors are multi-dimensional arrays of
numerical values. They are ubiquitous in the fields such as general relativity and
quantum chemistry. Tensors encountered in these fields usually
possess certain permutation symmetry. For instance, this happens to
the two-electron integrals and amplitudes of excited configurations in electronic
structure models\cite{QCtrad}. A general $n$-tuple excitation from a reference
$|\Phi_0\rangle$ can be expressed as
$|\Psi_{\text{n-tuple}}\rangle=\sum_{\{p_i\},\{h_i\}}t^{p_1p_2\cdots p_n}_{h_1h_2\cdots
h_n}a^{p_1p_2\cdots p_n}_{h_1h_2\cdots h_n}|\Phi_0\rangle$, where $t^{p_1p_2\cdots p_n}_{h_1h_2\cdots h_n}$ is a $2n$-way
tensor and both indices $p_1p_2\cdots p_n$ and $h_1h_2\cdots h_n$ are antisymmetric with respect to
transpositions, respectively. Complicated equations involving tensor product expressions frequently appear in sophisticated theories such as the high-order coupled cluster theory.
In such cases, manual manipulations of the resulted equations become difficult and automatic derivations need to be used.
A key step in such automatic derivations is the collection of tensor product terms
that are identical by using permutation symmetries of tensors or
relabeling dummy indices. In this paper, we examine fundamental questions
about the permutation symmetry of tensor products, which will help to
solve this issue. Specifically, let a general tensor product be denoted by the form
\begin{eqnarray}
Z^E\triangleq X_{1}^{E_{1}I_{1}}X_{2}^{E_{2}I_{2}}\cdots
X_{k}^{E_{k}I_{k}},\label{tensorProduct}
\end{eqnarray}
where the tensor $\bm{X}_{i}$ represents the $i$-th factor, $E\triangleq E_1\cup
E_{2}\cup\cdots E_{k}$ form a disjoint partition of an external
index set $E$, $I_{i}$ is the set of internal indices that will be
contracted, and hence $I_{i}\cap I_{j}$ represent the set of
internal indices to be contracted between $\bm{X}_{i}$ and
$\bm{X}_{j}$ or more explicitly $X_i^{E_i I_i}$ and
$X_{j}^{E_{j}I_{j}}$ in the component form. A simple example
is $Z^{ab}_{ij}=\bar{g}^{ab}_{cd}t^{cd}_{ij}$, which appears as
one of the contribution to the coupled cluster amplitude equations.
Here, $\bar{g}^{ab}_{cd}$ is the antisymmetrized two-electron integrals
(antisymmetric in permuting $a,b$ or $c,d$), $t^{cd}_{ij}$ is the double excitation amplitude (antisymmetric in permuting $a,b$ or $i,j$),
$E=E_1\cup E_2=\{a,b,i,j\}$, $I_1=I_2=\{c,d\}$,
and the Einstein summation convention for repeated indices has been used.
The same term can be written in many equivalent ways, e.g.,
$Z^{ab}_{ij}=-\bar{g}^{ab}_{dc}t^{cd}_{ij}$ using the antisymmetry of $\bar{g}$
or $Z^{ab}_{ij}=\bar{g}^{ab}_{dc}t^{dc}_{ij}$, which can be viewed
as via either relabelling the dummy indices or using antisymmetry of
$\bar{g}^{ab}_{cd}$ and $t^{cd}_{ij}$ simultaneously. All these different tensor
product expressions correspond to the same final tensor, and identifying
their equivalence is a nontrivial problem in computer-based manipulations.

In general, suppose the tensors
$\bm{X}_{i}$ in Eq. \eqref{tensorProduct} have certain permutation symmetries,
then the following questions are of fundamental importance, because they will be faced in designing any general automatic derivation and simplification tools based on tensor product expressions.
\begin{description}
\item [Q1 (classification)] Whether two given tensor expressions of the form
\eqref{tensorProduct} are identical? If not, to what extent they are
different?

\item [Q2 (enumeration)] How many different tensors can be formed from
a same set of factors? For a given member, how many of products are
equivalent (numerically identical) to it?

\item [Q3 (canonical form/represenative)] Among the equivalent tensor expressions,
is it possible to select a unique representative?

\item [Q4 (permutation symmetry)] How much the permutation symmetries in the factors
are inherited by the contracted result $Z^E$? Or more mathematically,
what is the permutation symmetry group of $Z^E$?
\end{description}
Obviously, without permutation symmetry in the factors, all these questions are trivial.
However, in the presence of permutation symmetry, they can become quite involved,
since the possible algebraic forms of products in Eq. \eqref{tensorProduct}
can be enormous.

We briefly mention some historical developments of tools for tensor product expressions
in the field of quantum chemistry.
In automating the derivation of the high-spin open-shell coupled
cluster singles and doubles (CCSD), Janssen and
Schaefer\cite{janssen1991automated} defined a \emph{canonical form} for a tensor
product by the permuted index arrays obtained via all
possible index permutations of the factors $\bm{X}_i$.
In this way, two equivalent terms become identical and can be rapidly combined.
Such exhaustive procedure is complete (guaranteed to find the
optimum), but becomes impractical due to the \emph{exponential} scaling
for terms of form $(T_n)^p$ (power of $n$-fold excitation $T_n$)
with large $n$ or $p$.  The tensor contraction engine (TCE) developed later by Hirata et al.\cite{hirata2003tensor} followed the
similar idea, but in order to treat high-order tensors, in which the
above exhaustive permutations of all indices become formidable, some
kinds of sorting based on several predefined rules were used.
However, for some ''cyclic'' tensor contractions like $t^{p_1p_2}_{h_3h_4}
t^{p_5p_6}_{h_3h_4}t^{p_1p_2}_{p_5p_6}$, the simple sorting becomes ambiguous
as the sorting of tensors and the sorting of indices are intertwined,
and in this case the exhaustive permutations were used.
Therefore, more economic approaches need to be designed.
Such approaches should utilize the properties of the ''canonical'' functions, rather than blindly
searching all possibilities. A simple example is that for the cluster
amplitude $T_n$ only, if the canonical form for Q3 is defined by the ordering of
indices, then the problem is simply solved by a sorting procedure,
which can obviously be achieved with a cost at most $O(n^2)$ instead
of $O(n!)$. The definition of canonical form is in some sense
arbitrary, because it depends on some predefined criteria. However,
it should at least satisfies two conditions. First,
it should be universal which means it can be applied to any kind of
product in Eq. \eqref{tensorProduct}. Second, in order to make such
form of practical use, there should be efficient algorithms to
compute it. We noted that for Q3, there exists a widely used algorithm for index canonicalization,
usually referred as the Butler-Portugal algorithm\cite{portugal1999algorithmic,portugal2001group,manssur107032group,manssur2002group,manssur2004canon,martin2008xperm,niehoff2018faster},
based on finding single coset representatives for external (or free\cite{portugal2001group} as used in related papers) indices, and using Butler's double coset representative algorithm\cite{butler1984computing} for internal (or dummy\cite{manssur107032group}) indices. However, it is known to be exponential in the worst case, although improvements
for the special case with totally symmetric/antisymmetric tensors has been proposed very recently\cite{niehoff2018faster}.

In this work, we answered the questions Q1-Q4 using group and graph
theoretical approaches. In Sec. \ref{foundation}, we provide a
classification of all tensor product expressions by introducing five
equivalence relations. From the classification theory,
the ways for enumerating different tensor product expressions,
defining unambiguous canonical form, and computing
permutation symmetry group naturally emerge.
Thus, all the solutions for questions Q1-Q4 can be given.
However, one still needs efficient algorithms to compute the canonical form
defined for Q3 and the permutation symmetry group for $Z^E$ in Q4.
To this end, we develop a graph-based algorithm in Sec. \ref{algorithm}.
Inspired by diagrammatic methods in permutation theory\cite{paldus1975}
and tensor network diagrams\cite{chan2012low,orus2014practical},
a graphical representation of algebraic tensor product is first defined.
Then, by combining the classical backtrack search for permutation groups\cite{kreher1998combinatorial,seress2003permutation,holt2005handbook}
and the idea of partitions from graph isomorphism algorithms\cite{mckay2014practical,piperno2008search}, we proposed an efficient algorithm
to compute the canonical form of a tensor product and the
automorphism group of its associated graph. The latter can provide
detailed information about the permutation symmetry of the resulted
tensor. Compared with the existing algorithms,
our algorithm is more efficient, and in fact, is \emph{polynomial}
for the worst case in the Butler-Portugal algorithm.
At a more abstract level, while the conventional algorithm based on double set representative
is more algebraic, the present algorithm is more graphical, which leads to
a different framework for future improvements.
As we will show later, it is more natural and fruitful
to think about the canonicalization of tensor product expressions in terms of graphs.
Those readers who are only interested in Q3 can skip Sec. \ref{foundation} and
go directly to Sec. \ref{algorithm} for the graphical canonicalization algorithms, which is largely self-contained.
Sec. \ref{results} shows some results for several tensor
products encountered in quantum chemistry and an example
for the worse case in the Butler-Portugal algorithm.
The conclusion and outlook are presented in Sec. \ref{conclusion}.
A summary of the notations used in Sec. \ref{foundation}
and Sec. \ref{algorithm} is listed in Table \ref{tab:list}.

\begin{table}[ht]
\caption{List of concepts and notations.}\label{tab:list}\scriptsize
\begin{tabular}{cl}
\hline\hline
notation & explanation \\
\hline
\multicolumn{2}{c}{{\bf Group theoretical classification theory}} \\
$a\sim b$ & equivalence relation \\
$[x]$     & equivalent class \\
$\times$  & direct product \\
$\mathcal{G}(\bm{Z}) $ & permutation symmetry group of a tensor $\bm{Z}$ \\
$\mathcal{S}_n$ & symmetric group with degree $n$\\
$\grp(S)$ & symmetry group for the factor set $S=(\bm{X}_1,\cdots,\bm{X}_k)$ \\
$\Omega$  & integer set $\Omega=\{1,2,\cdots,D\}$ associated with $S$ \\
$\mathcal{I}$ & total index set $\mathcal{I}=E_1\cup I_1\cup \cdots \cup E_k\cup I_k$\\
$\pi(\Omega)$ & coloring of $\Omega$: $\pi(\Omega)=\{\pi(1),\cdots,\pi(D)\}$\\
$\Omega_{E,I}(\pi)$ & support for external or internal labels \\
$\bar{\pi}$ & $\grp(S)$-equivalent class $\{g\circ\pi(\Omega):g\in\grp(S)\}$ \\
$\bar{\bar{\pi}}$ & $(\grp(S),\mathcal{H})$-equivalent class $\{h\circ\bar{\pi}:h\in\subgrp\}$ \\
$\grp_{(\pi)}(S)$ & pointwise stabilizer $\{g\in\grp(S):g\circ\pi(i)=\pi(i),\forall
i\in\Omega\}$ \\
$\varpi(\pi)$ & contraction pattern $\{\theta(i,j):\pi^I(i)=\pi^I(j),\forall
i,j\in\Omega_I(\pi)\}$ \\
$\grp_{\varpi(\pi)}(S)$ & setwise stabilizer $\{g\in\grp(S):g\circ\varpi(\pi)=\varpi(\pi)\}$ \\
$A \geq B$ & subgroup relation: $B$ is a subgroup of $A$ \\
$A \vartriangleright B$ & normal subgroup relation \\
$\lcr$ & left coset representatives $\{g_r:r\in L\}$ \\
$\bar{\pi}=\bigcup_{r\in L}\Class{\pi_r}$ & decomposition of $\bar{\pi}$
based on $\grp(S)=\bigcup_{r\in L} g_r\grp_{\varpi(\pi)}(S)$ \\
$\Class{\pi_r}$ & equivalent class $\{g\circ\pi_r(\Omega):g\in\grp_{\varpi(\pi_r)}(S)\}$  \\
$\phi$ & homomorphism in Theorem \ref{homoDecomposition} \\
$\Ker\phi$ & kernel of $\phi$ \\
$\grp(\Omega_E(\pi))$ & permutation symmetry group of $Z^E$ \\
$\Class{\pi_r^E}$ & equivalent class $\{g\circ\pi_r^E(\Omega_E):g\in\grp(\Omega_E(\pi_r))\}$ \\
$\Class{\pi_r^I}$ & equivalent class $\{g\circ\pi^I_r(\Omega_I):g\in\Ker\phi\}$ \\
$\pi_{\canon}=\mathcal{C}_\pi$ & representative coloring (canonical form) of $\bar{\bar{\pi}}$ \\
\\
\multicolumn{2}{c}{{\bf Graphical canonicalization algorithms}}\\
$G=(\Vset,\Eset)$ & graphical representation with vertices $\Vset$ and edges $\Eset$ \\
$\Vset$ & vertex set $\Vset=\Omega$ \\
$\Eset$ & edge set $\{(1,j):j\in\Omega_E(\pi)\}\cup \{(i,j):i,j\in\Omega_I(\pi)\}$  \\
$\mathcal{C}(G)$ & canonical form of $G$ \\
$\Aut_{\grp(S)}(G^E)$ & automorphism group of externally labeled graph $G^E$ \\
$\langle g_1,\cdots,g_k\rangle$ & a group generated by generators $g_1,\cdots,g_k$.\\
$\grp^{[i]}$ & pointwise stabilizer of the $i-1$ first elements of $\Omega$ \\
$u_{k,i_k}$ & element in the representation $g=u_{1,i_1}\cdots u_{D,i_D}$ \\
$\mathcal{T}$ & search tree of partial images or partitions \\
$\Pi$ & partition of $\mathcal{V}$ \\
$\Pi^g$ & image of $\Pi$ under the action of permutation $g$ \\
$\Pi\downarrow v$ & individualization $(\Pi_1,\cdots,\{v\},\Pi_{i}\backslash\{v\},\cdots,\Pi_r)$  \\
$\mathcal{R}(G,\Pi)$ & refinement of a partition \\
\hline\hline
\end{tabular}
\end{table}

\section{Group theoretical classification theory}\label{foundation}
\subsection{Equivalence relations, symmetries, and colorings}\label{class}

The basic tool we used for classification is the concept of
equivalence relation.
\begin{dfn}[equivalence relation]
A given binary relation $\sim$ on a set $M$ is said to be an
equivalence relation if and only if it satisfies three
requirements: (1) (reflexivity) $a \sim a$, (2) (symmetry) if $a\sim
b$ then $b\sim a$, (3) (transitivity) if $a\sim b$ and $b\sim c$
then $a\sim c$. The equivalence relation partitions the set $M$ into
disjoint equivalence classes $[x]$, which are defined via
$[x]=\{x'\in M | x'\sim x\}$. For two equivalence relations $\sim$
and $\approx$ defined on the same set $X$, and $a\approx b$ implies
$a\sim b$ for all $a,b \in X$, then $\sim$ is said to be a coarser
relation than $\approx$, and $\approx$ is a finer relation than
$\sim$.
\end{dfn}

The permutation symmetry of a tensor can be characterized by its
associated permutation group defined in the following way.
\begin{dfn}[tensor symmetry]
For a $r$-way tensor $\bm{Z}$, the set of permutations satisfying
the condition
\begin{eqnarray}
g_i\circ Z^P \triangleq Z^{g_i(P)}= Z^P,\quad
g_i\in\mathcal{S}_r,\label{tensorsym}
\end{eqnarray}
where $\mathcal{S}_r$ is the symmetric group of degree $r$, and $P$
is a set of abstract indices,
\begin{eqnarray}
g_i(P)=g_i(p_1p_2\cdots p_r)=p_{g_i(1)}p_{g_i(2)}\cdots p_{g_i(r)}
\end{eqnarray}
forms a permutation group under the composition of permutations. We
refer it as the permutation symmetry group of the tensor $\bm{Z}$,
denoted by $\grp(\bm{Z})$.
\end{dfn}

The meaning of Eq. \eqref{tensorsym} is transparent. It reveals that
the elements of $\bm{Z}$ are not all independent, but related in some way
via permutation. It can be viewed
as an extension of the transpositional symmetry of matrix, in which
case we can have symmetry and antisymmetric matrices $(12)\circ
A_{p_1p_2}=A_{p_2p_1}=\pm A_{p_1p_2}$. Note that the antisymmetry is
not covered by the definition Eq. \eqref{tensorsym}. In principle, we can exploit
more symmetry in $\bm{Z}$, e.g., by considering the permutations
whose actions only change the phase of the tensors. Such cases can be
easily incorporated in our framework introduced below by defining more
general permutation symmetry groups. For simplicity, in
the following discussions, we only consider the
symmetry defined in Eq. \eqref{tensorsym}.

To answer the questions Q1-Q4,
we need to define them more precisely. Some mathematical definitions in
the following context can be quite formal, and in such cases
it is suggested to go to Sec. \ref{example} for concrete examples.

\begin{dfn}[symmetry equivalent tensor product expressions]\label{equivTensor}
Two tensor product expressions $\bm{Z}_1$ and $\bm{Z}_2$ of form
\eqref{tensorProduct} are symmetry equivalent, if there exist a
permutation $g$ of factors $\{\bm{X}_i\}$ and indices ($E$ and $I$) such that
$\bm{Z}_1=g\circ\bm{Z}_2$ after some necessary relabeling of
internal indices. In this case, we denote them by
$\bm{Z}_1\sim\bm{Z}_2$.
\end{dfn}

By this definition, the first necessary condition for
$\bm{Z}_1\sim\bm{Z}_2$ is that they must share the same set of
factors, otherwise, even if neglecting the indices it is not
possible to match them by rearranging the factors. Let us denote
this factor set by $S=\{\bm{X}_1,\bm{X}_2,\cdots,\bm{X}_k\}$.
It is always possible to define an order for all possible orderings of
$S$. For instance, we can use a simple lexicographical order,
namely, for two different orderings of $S$,
$s_1=(\bm{X}_1,\bm{X}_2,\cdots,\bm{X}_k)$ and
$s_1'=(\bm{X}'_1,\bm{X}'_2,\cdots,\bm{X}'_k)$, we say $s_1<s_1'$ if
and only if the first $\bm{X}_i$, which is different from
$\bm{X}'_i$, comes before $\bm{X}'_i$ in the alphabet. From now on,
we assume that $S$ has been ordered by a user defined ordering for
tensor factors.

\begin{dfn}[factor set and the associated permutation group]\label{def:factorSet}
Given the factor set $S=(\bm{X}_1,\bm{X}_2,\cdots,\bm{X}_k)$ and the
permutation symmetry group of each factor $\grp(\bm{X}_i)$, we can
define the symmetry group of $S$ via direct product, viz.,
$\grp(S)=\grp(\bm{X}_1)\times\grp(\bm{X}_2)\times\cdots\times\grp(\bm{X}_k)$.
If some factors are the same, e.g.,
$\bm{X}_{i}=\bm{X}_{i+1}=\cdots=\bm{X}_{i+n-1}$, then the
corresponding parts of direct product $\grp_n(\bm{X}_i)\triangleq
\grp(\bm{X}_{i})\times\grp(\bm{X}_{i})\times\cdots\times\grp(\bm{X}_{i})$
should be replaced by the semidirect product
$\grp_n(\bm{X}_i)\rtimes \mathcal{S}_n(\bm{X}_i)$, where
$\mathcal{S}_n(\bm{X}_i)$, which is isomorphic to the symmetric group $\mathcal{S}_n$, represents
the symmetric group for permutations of the identical factors.
\end{dfn}

We can label each dimension of $S$ by a consecutive integer number.
The action of $\grp(S)$ on $S$ naturally induces an action on the
integer set $\Omega=\{1,2,\cdots,D\}$ where $D=\sum_{i=1}^k d_i$
with $d_i$ being the dimension of the tensor $\bm{X}_i$. For
simplicity, since $\grp(S)$ of $S$ and the induced permutation group of
$\Omega$ are isomorphic, we do not distinguish them and denote both
groups by $\grp(S)$. Now we focus on the index structure of tensor
products.
The indices $\pi(\Omega)\triangleq\{\pi(1),\pi(2),\cdots,\pi(D)\}$
extracted from \eqref{tensorProduct}, where the value of $\pi(i)$ belongs to
the total index set $\mathcal{I}=E_1\cup I_1\cup \cdots \cup E_k\cup I_k$,
can be viewed as a \emph{coloring} of $\Omega$, viz., a
mapping from $\Omega$ to a color set $\mathcal{I}$.
Then, we can establish the following connection.

\begin{thm}[tensor products and colorings]
The classification of different tensor products
\eqref{tensorProduct} with the same factor set $S$ under the
permutation symmetry group $\grp(S)$ is equivalent to the
classification of different colorings $\pi$ of $\Omega$ under the
induced permutation group on $\Omega$.
\end{thm}

In this work, we are interested in the tensor products \eqref{tensorProduct}
in which each internal index appear only twice. This is usually the case as
required by the invariance of equations under (orbital) rotations.
Because the internal indices in tensor products are free
to be permuted (relabeled) without changing the final result, we
have to introduce a group $\subgrp$ to describe the invariance for
permutation of colorings $\mathcal{I}$. The group $\subgrp$ is determined from the
types of internal indices (e.g., occupied or virtual orbitals
in quantum chemistry), and only the
internal indices of the same type are allowed to be permuted. For
a given number of external indices $n_{ext}$ and contracted internal
pairs $n_{c}$ for Eq. \eqref{tensorProduct}, we have $n_{int}=2n_{c}$, $D=n_{ext}+n_{int}$, and the
number of colorings $|\mathcal{I}|=n_{ext}+n_{c}$. It is easy to see that
the total number of different tensor product expressions of form \eqref{tensorProduct}
is given by
\begin{eqnarray}
N(n_{ext},n_{int})=D!/2^{n_c}.
\end{eqnarray}
However, as mentioned before, most of them correspond to the same
final result when considering the actions of $\grp(S)$ and $\subgrp$.
This point is formalized by the following equivalence
relation.

\begin{dfn}[equivalence of colorings]
Two colorings $\pi_1$ and $\pi_2$ are equivalent if and only if
there exist $g\in\grp(S)$ and $h\in\subgrp$ such that $\pi_1
g=h\pi_2$, defined under the composition $\pi_1(g(i))=h(\pi_2(i))$
where $i,g(i)\in\Omega$ and $\pi(i),h(\pi(i))\in\mathcal{I}$.
For simplicity, we refer this equivalence relation as
$(\grp(S),\subgrp)$-equivalence.
\end{dfn}

This definition essentially characterizes the Definition
\ref{equivTensor} for tensor products \eqref{tensorProduct}
in a more abstract way. Under this equivalence relation, the $N(n_{ext},n_{int})$ different
expressions can be classified into equivalent classes,
such that different expressions within each class correspond to
the same final result. The enumeration of nonequivalent colorings in the presence of
permutation symmetries $\grp(S)$ and $\subgrp$ is a classical
combinatorial problem that is solved by de Bruijn's
generalization\cite{de1967color,de1971survey} of the P\'{o}lya-Redfield
theory\cite{polya1937kombinatorische,redfield1927theory}.

\begin{lem}[de Bruijn's enumeration formula - basis for Q2]\label{counting}
Suppose $\mathcal{Y}=\{y_1,\cdots,y_{|\mathcal{I}|}\}$ is a set of colors, and
$\subgrp$ is a subgroup of the symmetric group $\mathcal{S}_{|\mathcal{I}|}$, then the generating function for
the colorings of $\Omega$ which are nonequivalent with respect to
the action of $\grp(S)$ on $\Omega$ and the action of $\subgrp$ on
$\mathcal{Y}$ can be obtained by identifying equivalent color
patterns in the polynomial
\begin{eqnarray}
F_{\grp,\subgrp}(y)=\frac{1}{|\subgrp|}\sum_{h\in\subgrp}P_{\grp}(\alpha_1(h),\alpha_2(h),\cdots,\alpha_D(h))\label{pattern}
\end{eqnarray}
where $P_{\grp}(x_1,x_2,\cdots,x_D)$ is the cycle index of $\grp$
defined by
\begin{eqnarray}
P_{\grp}(x_1,x_2,\cdots,x_D)=\frac{1}{|\grp|}\sum_{g\in\grp}\prod_{i=1}^n
x_{l_i}
\end{eqnarray}
with $g$ being a product of $n$ cycles, and the $i$-th cycle has
length $l_i$. The $\alpha_m(h)$ is defined by
\begin{eqnarray}
\alpha_m(h)=\sum_{\{j\;:\;h^m(j)=j,1\le j\le
|\mathcal{I}|\}}\prod_{i=0}^{m-1}y_{h^i(j)},
\end{eqnarray}
for $1\leq m\leq D$.
\end{lem}
This lemma can be used as the basis for answering Q2.
Suppose the first $n_{ext}$ elements of $\mathcal{Y}$ correspond to
external indices, while the remaining elements correspond to internal
indices, by using Lemma \ref{counting} one can find the
number of nonequivalent tensor products
from the coefficient of the monomial $y_1y_2\cdots
y_{n_{ext}}y_{n_{ext}+1}^2y_{n_{ext}+2}^2\cdots y_{|\mathcal{I}|}^2$. By setting
$\subgrp=\{e\}$, Lemma \ref{counting} reduces to the P\'{o}lya's
theorem\cite{polya1937kombinatorische} and then Eq. \eqref{pattern} gives the number of
nonequivalent classes under the equivalence relation $\pi_1=\pi_2 g$
($g\in\grp(S))$ for $\pi_1$ and $\pi_2$. We refer this
equivalence relation as $\grp(S)$-equivalence. For a coloring $\pi$,
its $\grp(S)$-equivalent class is denoted by
\begin{eqnarray}
\bar{\pi}=\{g\circ\pi(\Omega):g\in\grp(S)\},
\end{eqnarray}
with the induced action defined by $g\circ\pi(i)=\pi(g(i))$ for $g\in\grp(S)$ and $i\in\Omega$,
and the $(\grp(S),\subgrp)$-equivalent class is denoted by
\begin{eqnarray}
\bar{\bar{\pi}}=\{h\circ\bar{\pi}:h\in\subgrp\},
\end{eqnarray}
where we used the same notation $\circ$ for the action of $h$.
Clearly, $\grp(S)$-equivalence is finer than the $(\grp(S),\subgrp)$-equivalence,
since if $\pi_1$ and $\pi_2$ are $\grp(S)$-equivalent meaning that
they can be related by a permutation in $\grp(S)$, then
they obviously belong to the same $(\grp(S),\subgrp)$-equivalent class.
[NB: Here, we draw a connection with the double coset
based approach\cite{portugal1999algorithmic,portugal2001group,manssur107032group,manssur2002group,manssur2004canon,martin2008xperm}.
While we will focus on the classification of colorings $\pi$, the double coset based approach
focus the permutations on $\Omega$. Our $\grp(S)$ corresponds to the slot-symmetry group $S$,
$\subgrp$ permuting colors is isomorphic to the index-symmetry group $D$,
and $\bar{\bar{\pi}}$ is the counterpart of the double
coset $S\cdot g\cdot D$\cite{manssur2002group,martin2008xperm}.
However, as will be shown below and in the section for canonicalization algorithm,
in our case essentially the group $\subgrp$ does not need to be used explicitly.]

\subsection{Classification of symmetry equivalent tensor product expressions via a group chain}
All the symmetry equivalent tensor product expressions, whose corresponding colorings belong to the same $(\grp(S),\subgrp)$-equivalent class, correspond to the same final
tensor. For the purpose of eventually defining an unambitious canonical form (Q3), we need to further distinguish them.
To this end, we do not need to consider $\bar{\bar{\pi}}$, but just to focus on the classification of the finer class $\bar{\pi}$ (vide post).
The number of colorings in $\bar{\pi}$ is given by
\begin{eqnarray}
|\bar{\pi}|=|\grp(S)|/|\grp_{(\pi)}(S)|,\label{numOfExpr}
\end{eqnarray}
where the group $\grp_{(\pi)}(S)$ is the \emph{pointwise} stabilizer of
the coloring $\pi$ in $\grp(S)$, viz.,
\begin{eqnarray}
\grp_{(\pi)}(S)=\{g\in\grp(S):g\circ\pi(i)=\pi(i),\forall
i\in\Omega\}.
\end{eqnarray}
To further classify the $|\bar{\pi}|$ different colorings,
we introduce the notation of the color/contraction patterns for internal
indices as explained below: For a given $\pi$, we can rewrite it as $\pi=\pi^E\cup\pi^I$, which distinguishes the parts corresponding to external and internal indices. Consequently, $\Omega$ for given $\pi$ can be partitioned into a disjoint
union of $\Omega_E(\pi)$ and $\Omega_I(\pi)$, which are supports of
$\pi^E$ and $\pi^I$, respectively. Then, we can construct a set of
unordered pairs by
\begin{eqnarray}
\varpi(\pi)=\{\theta(i,j):\pi^I(i)=\pi^I(j),\forall
i,j\in\Omega_I(\pi)\},
\end{eqnarray}
where the head $\theta$ is used to distinguish different types of
internal indices. If all the types of internal indices are the same,
the $\theta(i,j)=\{i,j\}$ can just be a set.
In terms of tensor products, we can call
$\varpi(\pi)$ as the \emph{contraction pattern} of $\pi$. We say
$\pi_1,\pi_2\in\bar{\pi}$ are equivalent, if
$\varpi(\pi_1)=\varpi(\pi_2)$. It is easy to verify that this is
indeed an equivalence relation on $\bar{\pi}$, with
the induced action of $g$ on $\varpi(\pi)$ can be defined as
$g\circ\theta(i,j)=\theta(g(i),g(j))$. It deserves to
be emphasized again that $\theta(g(i),g(j))$ is an unordered pair.

Given $\varpi(\pi)$, its \emph{setwise} stabilizer is denoted by
\begin{eqnarray}
\grp_{\varpi(\pi)}(S)=\{g\in\grp(S):g\circ\varpi(\pi)=\varpi(\pi)\}.
\end{eqnarray}
Then, it is important to realize the group chain relation
\begin{widetext}
\begin{eqnarray}
\grp(S) \geq \grp_{\Omega_I(\pi)}(S)=\grp_{\Omega_E(\pi)}(S)
\geq \grp_{\varpi(\pi)}(S) \vartriangleright
\left(
\grp_{\varpi(\pi)}(S)\cap\grp_{(\Omega_E(\pi))}(S)
\right)
\vartriangleright
\grp_{(\pi)}(S),\label{chain}
\end{eqnarray}
\end{widetext}
where $\geq$ and $\vartriangleright$ represent subgroup relation and
normal subgroup relation, respectively. The subgroups
$\grp_{\Omega_I(\pi)}(S)$ and $\grp_{\Omega_E(\pi)}(S)$
represent the setwise stabilizers of $\Omega_I(\pi)$ and
$\Omega_E(\pi)$, respectively.
They are simply the same, because $\forall
g\in\grp(S)$, $g(i)\in \Omega_I(\pi),\forall i\in\Omega_I(\pi)$ is
equivalent to say $g(i)\in \Omega_E(\pi),\forall i\in\Omega_E(\pi)$. Eq.
\eqref{chain} shows that $\grp_{\varpi(\pi)}(S)$ is a subgroup of
$\grp_{\Omega_I(\pi)}(S)$.
The group $\grp_{\varpi(\pi)}(S)\cap\grp_{(\Omega_E(\pi))}(S)$ is the pointwise stabilizer of $\Omega_E(\pi)$ (or equivalently $\pi^E$ via the induced action
$g\circ\pi^E(i)=\pi^E(g(i))$) in $\grp_{\varpi(\pi)}(S)$.
Finally, the group $\grp_{(\pi)}(S)$ is a normal subgroup of $\grp_{\varpi(\pi)}(S)\cap\grp_{(\Omega_E(\pi))}(S)$.
The importance of Eq. \eqref{chain} is that it allows to further classify
the colorings in $\bar{\pi}$ according to the group chain using
coset decomposition.

The first level decomposition $\grp(S)
\geq \grp_{\Omega_I(\pi)}(S)=\grp_{\Omega_E(\pi)}(S)$
classifies $\bar{\pi}$ into $|\grp(S)|/|\grp_{\Omega_E(\pi)}(S)|$
classes, such that different classes have the different $\Omega_E(\pi)$.
The next level decomposition
$\grp_{\Omega_I(\pi)}(S)=\grp_{\Omega_E(\pi)}(S)\geq
\grp_{\varpi(\pi)}(S)$ classifies the colorings with the same
$\Omega_E(\pi)$ according to their contraction pattern. In total,
the nonequivalent classes with respect to the mapping $\varpi$ can
be obtained from the left coset decomposition
\begin{eqnarray}
\grp(S)=\bigcup_{r\in L} g_r\grp_{\varpi(\pi)}(S),\label{lcr}
\end{eqnarray}
where $L$ is an index set and the set of left coset representatives
$g_r$ is denoted by $\lcr=\{g_r:r\in L\}$. Eq. \eqref{lcr} induces a
partition of $\bar{\pi}$ into disjoint classes with the same
cardinality,
\begin{eqnarray}
\bar{\pi}&=&\bigcup_{r\in L}\Class{\pi_r} ,\quad
\pi_r=g_r\circ\pi,\nonumber\\
\Class{\pi_r}&=&\{g\circ\pi_r(\Omega):g\in\grp_{\varpi(\pi_r)}(S)\},\label{patternDecomposition}
\end{eqnarray}
where the class $\Class{\pi}$, i.e., the orbit of $\pi$ under
$\grp_{\varpi(\pi)}(S)$, is the equivalent class of $\pi$ under the
equivalence relation $\varpi(\pi_1)=\varpi(\pi_2)$.
Note that $\grp_{\varpi(\pi_r)}(S)=g_r\grp_{\varpi(\pi)}(S)g_r^{-1}$, because
for $g\in\grp_{\varpi(\pi)}(S)$, we have $g_r g
g_r^{-1}\circ\varpi(\pi_r)=g_r g
g_r^{-1}g_r\circ\varpi(\pi)=g_r\circ\varpi(\pi)=\varpi(\pi_r)$. According
to Eq. \eqref{lcr}, the number of nonequivalent classes $\Class{\pi_r}$ is
\begin{eqnarray}
|\lcr|=|\grp(S)|/|\grp_{\varpi(\pi)}(S)|.\label{numOfSkeleton}
\end{eqnarray}
The cardinality of the class $\Class{\pi_r}$ is given by
\begin{eqnarray}
|\Class{\pi_r}|=|\grp_{\varpi(\pi)}(S)|/|\grp_{(\pi)}(S)|,\quad\forall
r\in L,\label{numOfMember}
\end{eqnarray}
which is the order of the quotient group
\begin{eqnarray}
\grp_{\varpi(\pi)}(S)/\grp_{(\pi)}(S)=\{g\grp_{(\pi)}(S): g\in
\grp_{\varpi(\pi)}(S)\}.
\end{eqnarray}
Note the relation $|\bar{\pi}|=|\lcr|\cdot|\Class{\pi_r}|$ is indeed
fulfilled by Eqs. \eqref{numOfExpr}, \eqref{numOfSkeleton}, and
\eqref{numOfMember}.

Next, since $\grp_{\varpi(\pi)}(S)$ is a subgroup of
$\grp_{\Omega_E(\pi)}(S)$, its action leaves $\Omega_E(\pi)$
invariant, and simply induces permutations on $\Omega_E(\pi)$. Let
us look into this induced action in details.

\begin{thm}[permutation symmetry group of $Z^E$ - answer for Q4]\label{homoDecomposition}
The mapping from $\grp_{\varpi(\pi)}(S)$ to a permutation group on
$\Omega_E(\pi)$ denoted by $\grp(\Omega_E(\pi))$ defined via
\begin{eqnarray}
\phi:\grp_{\varpi(\pi)}(S)&\longrightarrow&\grp(\Omega_E(\pi))\nonumber\\
g&\longmapsto&\phi_g: \phi_g(i)\triangleq g(i),\forall
i\in\Omega_E(\pi),\label{Homo}
\end{eqnarray}
is a group homomorphism. The kernel of the mapping
$\Ker\phi=\{g\in\grp_{\varpi(\pi)}(S):\phi_g=e\}=
\grp_{\varpi(\pi)}\cap\grp_{(\Omega_E(\pi))}(S)$ is the pointwise stabilizer of $\Omega_E(\pi)$ in $\grp_{\varpi(\pi)}(S)$.
The group $\grp(\Omega_E(\pi))$,
isomorphic to the quotient group $\grp_{\varpi(\pi)}(S)/\Ker\phi$,
gives the permutation symmetry group of $Z^E$.
\end{thm}

The proof is straightforward. By noting $\forall i\in
\Omega_E(\pi)$,
$\phi_{g_1g_2}(i)=g_{1}g_{2}(i)=g_{1}(g_{2}(i))=\phi_{g_1}\phi_{g_2}(i)$,
thus $\phi_{g_1g_2}=\phi_{g_1}\phi_{g_2}$. Besides, $\phi_e=e$ and
$\phi_g^{-1}=\phi_{g^{-1}}$. Therefore, the so-constructed
$\grp(\Omega_E(\pi))$ indeed form a permutation group acting on
$\Omega_E(\pi)$, and $\phi$ is a group homomorphism. According to
the first isomorphism theorem of groups, $\Ker\phi$ is a normal
subgroup of $\grp_{\varpi(\pi)}(S)$, and $\grp(\Omega_E(\pi))$ is
isomorphic to the quotient group $\grp_{\varpi(\pi)}(S)/\Ker\phi$.
Note that the group $\Ker\phi$ can be considered as effectively acting on $\Omega_I$ only.

For other contraction patterns $\varpi(\pi_r)$ in Eq.
\eqref{patternDecomposition}, Eq. \eqref{Homo} leads to
$\grp(\Omega_E(\pi_r))\triangleq
\phi(\grp_{\varpi(\pi_r)}(S))=g_r\grp(\Omega_E(\pi))g_r^{-1}$. In
particular, if $g_r\in \grp_{\Omega_E(\pi)}(S)$ then
$\grp(\Omega_E(\pi_r))=\grp(\Omega_E(\pi))$. In sum, different
$\grp(\Omega_E(\pi_r))$ are isomorphic and if we relabel
$\Omega_E(\pi_r)$ by the same set of colors, i.e., the same set of
external indices, then these groups will induce exactly the same group on
the colors (external indices). This is in fact the permutation
symmetry group $Z^E$ required in Q4.

With this theorem, we can further partition $\Class{\pi_r}$ in Eq.
\eqref{patternDecomposition} as a direct product
\begin{eqnarray}
\Class{\pi_r}&=&\Class{\pi_r^E}\times\Class{\pi_r^I},\label{classDecomp}\\
\Class{\pi_r^E}&=&\{g\circ\pi_r^E(\Omega_E):g\in\grp(\Omega_E(\pi_r))\},\label{classpPiE}\\
\Class{\pi_r^I}&=&\{g\circ\pi^I_r(\Omega_I):g\in\Ker\phi\}.\label{classPiI}
\end{eqnarray}
The meaning of decomposition is quite clear. Namely,
the external part $\Class{\pi_r^E}$ is composed of the images of
$\pi_r^E$ under the action of $\grp(\Omega_E(\pi_r)$,
while the internal part $\Class{\pi_r^I}$ is composed
of the images of $\pi^I$ under the action of $\Ker\phi$.
Thus, the cardinality of $\Class{\pi_r^E}$ is given by
\begin{eqnarray}
|\Class{\pi_r^E}|=|\grp(\Omega_E(\pi_r))|,\label{piE}
\end{eqnarray}
since the colors in $\pi_r^E$ are all different, while the
cardinality of $\Class{\pi_r^I}$ is given by
\begin{eqnarray}
|\Class{\pi_r^I}|=|\Ker\phi|/|\grp_{(\pi)}(S)|,\label{piI}
\end{eqnarray}
which is the order of the quotient group $\Ker\phi/\grp_{(\pi)}(S)$.
Note that $|\Class{\pi_r^E}|\cdot|\Class{\pi_r^I}|=|\Class{\pi_r}|$
indeed recovers Eq. \eqref{numOfMember}.

In summary, for $|\bar{\pi}|$ \eqref{numOfExpr}, due to the group chain \eqref{chain},
now we can have
\begin{eqnarray}
|\bar{\pi}|=|\mathcal{L}|\cdot|\Class{\pi_r^E}|\cdot|\Class{\pi_r^I}|.
\end{eqnarray}
Finally, we mention that while $\subgrp$ is not considered in this subsection,
the number of $\bar{\pi}$ classes that are equivalent with respect to $\subgrp$
can be found as
$|\bar{\bar{\pi}}|/|\bar{\pi}|=|\subgrp|/|\Ker\phi/\grp_{(\pi)}(S)|$.
By changing the colors for
internals (or equivalently, relabeling the internal indices), these
classes can be related, and they corresponds to the same final tensor.
For the later convenience, we introduce the following notation
\begin{eqnarray}
\overline{\Class{\pi_r^I}}=\{h\circ\pi^I_r(\Omega_I):h\in\subgrp\}.\label{HclassPiI}
\end{eqnarray}
for the enlarged class of $\Class{\pi_r^I}$ obtained by all possible
relabelings of the internal indices.

\subsection{Classification theory and canonical form}

Now we are able to answer Q1 for classification of tensor product expressions by the following theorem.
\begin{thm}[classification theory of tensor product expressions - answer for Q1]
All tensor product expressions can be
systematically classified based on the hierarchy:
\begin{enumerate}[(E1)]
\item\label{e1} the equivalence relation with respect to the factor set
$S$,

\item\label{e2} the equivalence relation with respect to both $\grp(S)$
and $\subgrp$ for colorings $\pi$,

\item\label{e3} the equivalence relation with respect to $\grp(S)$ only,

\item\label{e4} the equivalence relation with respect to $\grp_{\varpi(\pi)}(S)$ for the contraction pattern,

\item\label{e5} the equivalence relation with respect to $\grp(\Omega_E)$ for
$\pi^E$ and $\Ker\phi$ for $\pi^I$.
\end{enumerate}
The countings of nonequivalent classes for
(E\ref{e2}) can be performed based on de Bruijn's enumeration formula (Lemma
\ref{counting}), for (E\ref{e3}) based on Polya's theorem (Lemma
\ref{counting} with $\subgrp=\langle e\rangle$), for (E\ref{e4}) with
Eq. \eqref{lcr} for $|\lcr|$, for (E\ref{e5}) with Eqs. \eqref{piE} and
\eqref{piI} for $|\Class{\pi_r^E}|$ and $|\Class{\pi_r^I}$,
respectively.
\end{thm}
This theorem gives a way to check whether two tensor products \ref{tensorProduct} are identical theoretically. This means that by going through each step, one can see at which level two tensor products are different. However, given two expressions,
while (E\ref{e1}) is very easy to check, checking all other fours
directly may encounter \emph{exponential} complexity in the number of indices for large
$\grp(S)$ and $\subgrp$.

One way to simplify the comparison is based on defining a canonical form
(or representative). Then, before comparing two expressions,
they can be first transformed into their respective canonical forms,
and then if two canonical forms are different expressions, the original
terms are different. Based on the procedure in the above theorem,
we can provide a unambiguous definition of canonical form
for tensor product expressions, which can uniquely
select a representative for $\pi$ from its $(\grp(S),\subgrp)$-equivalent class $\bar{\bar{\pi}}$. This basically follows the group chain
\eqref{chain} and defines one representative for each step.

\begin{thm}[canonical form/representative - answer for Q3]\label{representative}
Within a $(\grp(S),\subgrp)$-equivalent class $\bar{\bar{\pi}}$, we
assume a priority of external indices over internal indices, then
the following four conditions uniquely define a coloring
$\pi_{\canon}=\mathcal{C}_\pi$ that
constitutes a representative (canonical form) of $\bar{\bar{\pi}}$,
in the sense (1) $\pi_{\canon}$ is $(\grp(S),\subgrp)$-equivalent to
$\pi$, (2) $\forall g\in\grp(S),h\in\subgrp$,
$\mathcal{C}_{h\pi}(g\Omega)=\mathcal{C}_\pi(\Omega)$.
\begin{enumerate}[(C1)]
\item\label{c1} The support $\Omega_E(\pi_{\canon})$ is minimal in lexicographical order
among $\{\Omega_E(\pi_r):r\in L\}$, which is equivalent to
$\Omega_I(\pi_{\canon})$ is maximal in lexicographical order among
$\{\Omega_I(\pi_r):r\in L\}$. This step will pick several $\Class{\pi_r}$ classes having the same minimal $\Omega_E(\pi_{\canon})$.

\item\label{c2} The contraction pattern $\varpi(\pi^I_{\canon})$ is minimal in lexicographical order among the colorings satisfying (C\ref{c1}), which will uniquely
    pick one $\Class{\pi_r}$ class.

\item\label{c3} $\pi_{\canon}^E$ is minimal in lexicographical order
 among all external colors in $\Class{\pi_r^E}$ derived from the decomposition $\Class{\pi_r}$ \eqref{classDecomp}, which will fix the external indices.

\item\label{c4} $\pi^I_{\canon}$ is minimal in lexicographical order among all internal colors in $\overline{\Class{\pi_r^I}}$, which will fix the internal indices.
\end{enumerate}
\end{thm}
It is important to note that in the relabelling step (C\ref{c4}) for internal indices, rather than finding the minimal $\pi^I_{\canon}$ in $\Class{\pi_r^I}$, the search has been extended to $\overline{\Class{\pi_r^I}}$, in order to take into account $\subgrp$ such that the final canonical form $\pi_{\canon}$ is for the whole class $\bar{\bar{\pi}}$ rather than only for $\bar{\pi}$. There are two special cases of this theorem.
\begin{enumerate}[(S1)]
\item\label{S1} If there is no external index, then only (C\ref{c2}) and
(C\ref{c4}) apply, because $\Omega_I(\pi)=\Omega$,
$\Omega_E(\pi)=\emptyset$ for any $\pi$.

\item\label{S2} If there is no internal index at all, then only (C\ref{c3})
applies, in which case $\varpi(\pi)=\emptyset$,
$\grp_{\varpi(\pi)}(S)=\grp(S)=\grp(\Omega_E(\pi))$, and
$\Ker\phi=\langle e\rangle$, because $\Omega_E(\pi)=\Omega$,
$\Omega_I(\pi)=\emptyset$ for any $\pi$.
\end{enumerate}

While Theorem \ref{representative} provides a well-defined
representative, in practice we still need an efficient algorithm to
compute it. Besides, we should emphasize that the canonical form defined by Theorem \ref{representative}
is not the only way to define canonical forms. As long as the way
to pick representatives in each step is well-defined, a
unique canonical form can be defined. In conjunction with
the freedom in defining ordering for factors in $S$
mentioned before Definition \ref{def:factorSet},
these freedoms in defining canonical forms may be utilized
to design efficient algorithms. We leave the study
of alternative definitions of canonical forms in future.
Before we step into the algorithm
for the calculation of representatives in Sec. \ref{algorithm}, it is
better to illustrate the above abstract results with some
concrete examples.

\subsection{Examples}\label{example}
\begin{figure}\centering
\resizebox{0.4\textwidth}{!}{\includegraphics{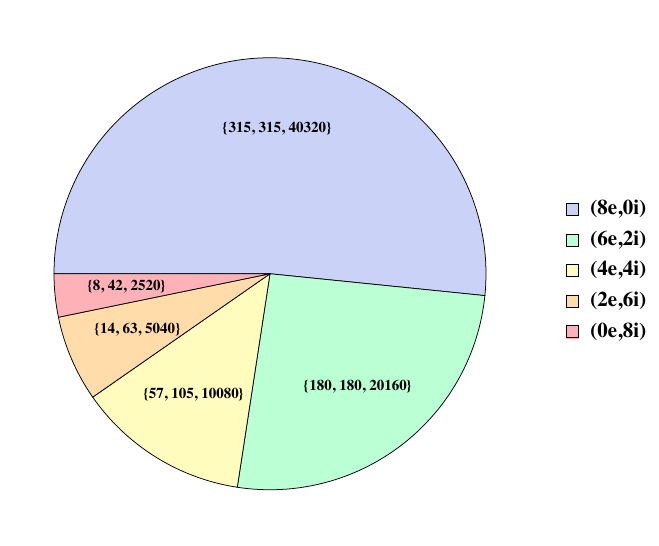}}
\caption{Classification of different tensor expressions sharing the
same factor set $S=(\bm{g},\bm{g})$ with
$\grp(\bm{g})=\langle(12),(34),(13)(24)\rangle$. The pair
$(n_{ext}\mathbf{e},n_{int}\mathbf{i})$ represents the coloring type
with $n_{ext}$ external indices and $n_{int}$ internal indices, The
triple $\{N_{gh},N_{g},N(n_{ext},n_{int})\}$ is a collection of
three numbers: $N_{gh}$ the number of $(\grp(S),\subgrp)$-equivalent
classes, $N_g$ the number of $\grp(S)$-equivalent classes, and
$N(n_{ext},n_{int})$ the total number of different
expressions.}\label{classification}
\end{figure}

To better illustrate the concepts introduced in the previous
section, we will consider simple tensor products formed by two $\bm{g}$
tensors, where the two electron integral tensor $g_{pq,rs}=[pq|rs]$
in the Mulliken notation\cite{QCtrad} satisfies the symmetry relation
\begin{eqnarray}
g_{pq,rs}=g_{qp,rs}=g_{pq,sr}=g_{rs,pq},
\end{eqnarray}
that is $\grp(\bm{g})=\langle(12),(34),(13)(24)\rangle$ with only
the generators for $\grp(\bm{g})$ listed in a cycle notation explicitly. The
symmetry group for the factor set $S=(\bm{g},\bm{g})$ is $\grp(S)=
(\grp(\bm{g})\times\grp(\bm{g}))\rtimes\mathcal{S}_2(\bm{g})
=\langle
(12),(34),(13)(24),(56),(78),(57)(68),(15)(26)(37)(48)\rangle$,
whose order is $8\times8\times2=128$. For the coloring type denoted
by $(n_{ext}\mathbf{e},n_{int}\mathbf{i})$ ($n_{ext}$ external
indices and $n_{int}$ internal indices), we have
$n_{ext}+n_{int}=D=8$ and $n_{c}=n_{int}/2$.
For simplicity, we will assume that all the internal indices are of
the same type such that $\subgrp=\mathcal{S}_{n_c}$ (the symmetric group of
degree $n_c$), and also the external indices are of the same type.
Then the number of $(\grp(S),\subgrp)$-equivalent classes $N_{gh}$
and the number of $\grp(S)$-equivalent classes $N_g$ can be
calculated from Lemma \ref{counting}. The results together with the
number of different expressions $N(n_{ext},n_{int})=D!/2^{n_c}$
($D=8$ in this example) for each coloring type are summarized in
Figure \ref{classification}. There are in total
$\sum_{n_c=0}^{4}8!/2^{n_c}=78120$ different tensor expressions sharing the same
factor set.

\begin{figure}\centering
\resizebox{0.4\textwidth}{!}{\includegraphics{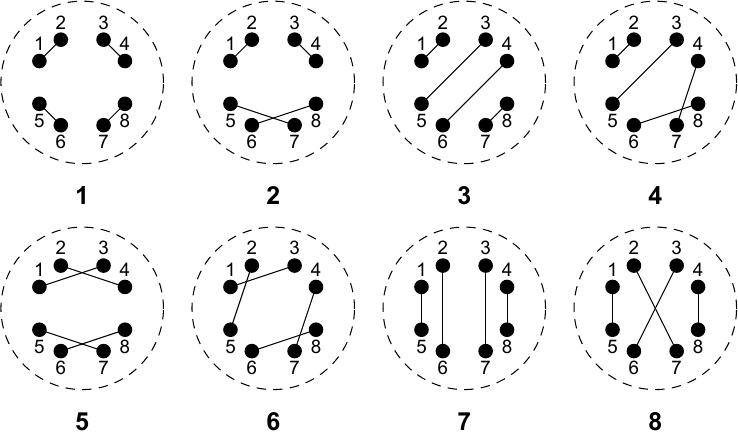}}
\caption{The representatives of
8 different groups of contraction patterns $\varpi(\pi)$
determined by the condition (C\ref{c2}).}\label{internalContractionPattern}
\end{figure}

We examine two special types first, either without internal indices or without
external indices. For the
$(8\mathbf{e},0\mathbf{i})$ type, there are
$8!=40320$ different expressions, e.g.,
$g_{e_1e_2,e_3e_4}g_{e_5e_6,e_7e_8}$,
$g_{e_1e_2,e_3e_4}g_{e_5e_6,e_8e_7}$, and
$g_{e_3e_2,e_1e_4}g_{e_5e_6,e_8e_7}$.
It is easy to see that they can be classified
into $N_{gh}=N_{g}=8!/128=315$ classes, since within each class
different expressions are simply
related by permutations in $\mathcal{G}(S)$.
For the $(0\mathbf{e},8\mathbf{i})$ type,
there are $8!/2^4=2520$ different expressions
such as $g_{i_1i_2,i_3i_4}g_{i_1i_2,i_3i_4}$
and $g_{i_4i_2,i_3i_1}g_{i_1i_2,i_3i_4}$.
By using permutations in $\mathcal{G}(S)$,
they can be classified into 42 $\mathcal{G}(S)$-equivalent classes.
These $\mathcal{G}(S)$-equivalent classes can further be
classified into 8 different groups of $(\mathcal{G}(S),\subgrp)$-equivalent classes.
The representatives determined by the condition (C\ref{c2})
for the contraction patterns $\varpi(\pi)$ of these 8 classes
are presented in Figure \ref{internalContractionPattern}.

\begin{figure}\centering
\resizebox{0.4\textwidth}{!}{\includegraphics{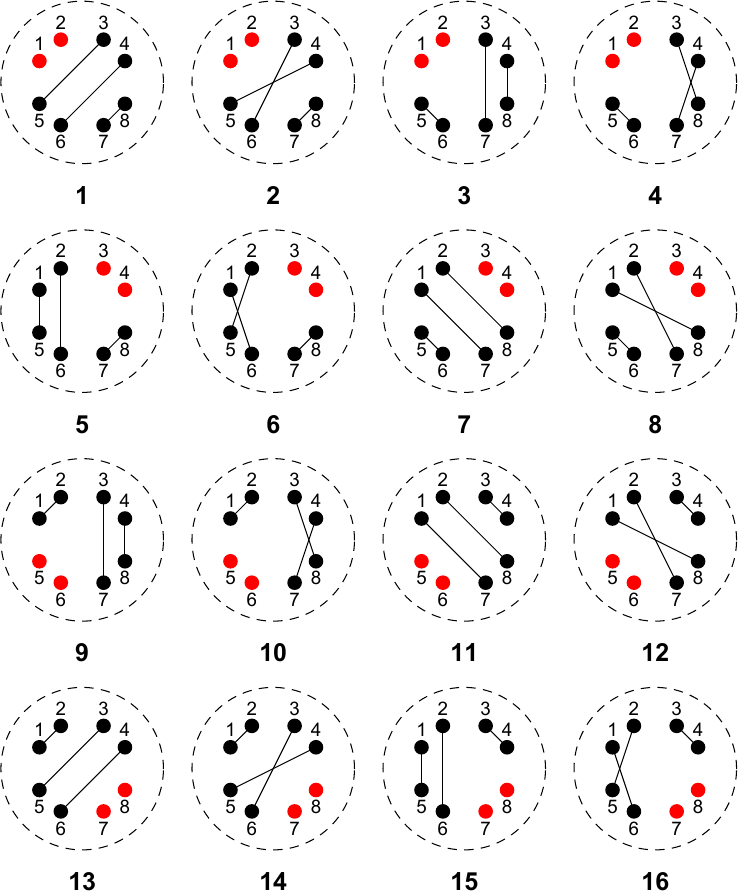}}
\caption{The 16 different contraction patterns $\varpi(\pi_r)$.
The initial coloring $\pi$ with $\varpi(\pi)$ given in Eq. \eqref{exampleGGpt}
belongs to the 15-th class. In canonicalization,
the condition (C\ref{c1}) for minimal $\Omega_{E}(\pi)$
and the condition (C\ref{c2}) for minimal
$\varpi(\pi)$ select the first pattern as the representative.
The mapping $g_{r}$ \eqref{patternDecomposition} from the 15-th to the first is given
in Eq. \eqref{gRmap}.}\label{contractionPattern}
\end{figure}

Next we consider an expression
$g_{i_1i_2,i_3i_3}g_{i_1i_2,e_1e_2}$,
which corresponds to the coloring $\pi=\{i_1, i_2, i_3, i_3, i_1, i_2, e_1, e_2\}$
to be canonicalized. Following the lines of Sec. \ref{class} we will investigate the
corresponding quantities $\bar{\bar{\pi}}$, $\bar{\pi}$,
$\Class{\pi}$, $\Class{\pi^E}$, $\Class{\pi^I}$, and the most
important one $\pi_{\canon}$. First, according to the definitions,
we have
\begin{eqnarray}
\pi^E&=&\{e_1,e_2\},\\
\Omega_E(\pi)&=&\{7,8\},\\
\pi^I&=&\{i_1, i_2, i_3, i_3, i_1, i_2\},\\
\Omega_I(\pi)&=&\{1,2,3,4,5,6\},\\
\varpi(\pi)&=&\{\theta(1,5),\theta(2,6),\theta(3,4)\}.\label{exampleGGpt}
\end{eqnarray}
and then the groups in Eq. \eqref{chain} are calculated as
\begin{eqnarray}
\grp_{\Omega_I(\pi)}(S)&=&\langle (12),(34),(56),(78),(13)(24)\rangle\nonumber\\
&=&\grp_{\Omega_E(\pi)}(S),\\
\grp_{\varpi(\pi)}(S)&=&\langle (34),(12)(56),(78) \rangle,\\
\grp_{(\pi)}(S)&=&\langle (34) \rangle,
\end{eqnarray}
with
\begin{eqnarray}
|\grp_{\Omega_I(\pi)}(S)|=32,\; |\grp_{\varpi(\pi)}(S)|=8,\;
|\grp_{(\pi)}(S)|=2.
\end{eqnarray}
Then, the number of expressions $|\bar{\pi}|$ \eqref{numOfExpr}, the
number of different contraction patterns $|\lcr|$
\eqref{numOfSkeleton}, and the number of colorings in any class
$\Class{\pi_r}$ \eqref{numOfMember} are found as
\begin{eqnarray}
|\bar{\pi}|&=&|\grp(S)|/|\grp_{(\pi)}(S)=128/2=64,\nonumber\\
|\lcr|&=&|\grp(S)|/|\grp_{\varpi(\pi)}(S)|=128/8=16,\nonumber\\
|\Class{\pi_r}|&=&|\grp_{\varpi(\pi)}(S)|/|\grp_{(\pi)}(S)|=8/2=4.
\end{eqnarray}
The 16 contraction patterns $\varpi(\pi_r)$ are illustrated pictorially in Figure
\ref{contractionPattern}, where the patterns in each row share the
same $\Omega_I(\pi_r)$. The 16 classes $\Class{\pi_r}$ can be divided
into $|\grp(S)|/|\grp_{\Omega_I(\pi)}(S)|=128/32=4$ groups based on
their supports $\Omega_I(\pi_r)$ (or equivalently $\Omega_E(\pi_r)$),
which correspond to 4 rows in Figure \ref{contractionPattern}.

The initial coloring $\pi$ with $\varpi(\pi)$ given in Eq. \eqref{exampleGGpt}
belongs to the 15-th class. In the selection of representatives for $\bar{\bar{\pi}}$,
the condition (C\ref{c1}) for minimal $\Omega_{E}(\pi)$
implies the contraction patterns in the first row of
Figure \ref{contractionPattern} should be selected,
while the condition (C\ref{c2}) for minimal
$\varpi(\pi)$ implies the first class is the choice,
because it is lexicographically smaller than the
other three class in the first row.
The permutation $g_{r}$ \eqref{patternDecomposition}
from the 15-th $\varpi(\pi)$ to the first contraction
pattern $\varpi(\pi_r)=\{\theta(3,5),\theta(4,6),\theta(7,8)\}$
can be found as
\begin{eqnarray}
g_r=\left(\begin{array}{cccccccc}
1 & 2 & 3 & 4 & 5 & 6 & 7 & 8 \\
5 & 6 & 7 & 8 & 3 & 4 & 1 & 2 \\
\end{array}\right),\label{gRmap}
\end{eqnarray}
in a two-line (image) notation for permutations.
For canonicalizing $\pi^E$ and $\pi^I$, according to Theorem
\ref{homoDecomposition}, we have
\begin{eqnarray}
\Ker\phi=\langle(34),(12)(56)\rangle,\;\;
\grp(\Omega_E(\pi))=\langle (78) \rangle,
\end{eqnarray}
such that
\begin{eqnarray}
\Class{\pi^E}&=&\{\{e_1,e_2\},\{e_2,e_1\}\},\nonumber\\
\Class{\pi^I}&=&\{\{i_1, i_2, i_3, i_3, i_1, i_2\}, \{i_2, i_1, i_3, i_3, i_2,i_1\}\}.
\end{eqnarray}
The condition (C\ref{c3}) selects $\pi_{\canon}^E=\{e_1,e_2\}$ from $\Class{\pi_r^E}$.
Using $g_{r}$ \eqref{gRmap} and $\Class{\pi^I}$,
the enlarged class $\overline{\Class{\pi_r^I}}$ can be found as
\begin{eqnarray}
\overline{\Class{\pi_r^I}}&=&
\{\{i_1, i_2, i_1, i_2, i_3, i_3\}, \{i_2, i_1, i_2, i_1, i_3, i_3\}, \nonumber\\
&&\;\;\{i_1, i_3, i_1, i_3, i_2, i_2\}, \{i_3, i_1, i_3, i_1, i_2, i_2\}, \nonumber\\
&&\;\;\{i_2, i_3, i_2, i_3, i_1, i_1\}, \{i_3, i_2, i_3, i_2, i_1, i_1\} \}.
\end{eqnarray}
It can be verified that $|\overline{\Class{\pi_r^I}}|/
|\Class{\pi^I}|=|\subgrp|/|\Ker\phi/\grp_{(\pi)}(S)|=3!/(4/2)=3$ gives the
number of $\bar{\pi}$ classes that are nonequivalent with respect to $\grp(S)$
but equivalent with respect to $\subgrp$.
From $\overline{\Class{\pi_r^I}}$, the condition (C\ref{c4}) suggests $\pi_{\canon}^I=\{i_1,i_2,i_1,i_2,i_3,i_3\}$, which can also be
simply found by a relabeling of $\varpi(\pi_r)=\{\theta(3,4),\theta(5,6),\theta(7,8)\}$.
Thus, we have $\pi_{\canon}=\{e_1,e_2,i_1,i_2,i_1,i_2,i_3,i_3\}$ and hence the
corresponding canonical form for
the input expression $g_{i_1i_2,i_3i_3}g_{i_1i_2,e_1e_2}$ is
$g_{e_1e_2,i_1i_2}g_{i_1i_2,i_3i_3}$.

Finally, it deserves to be emphasized that generally
$\grp(\Omega_E(\pi))$ is not a subgroup of $\grp(S)$. The following
example illustrates this fact. For $Z^{e_1e_2}\triangleq t_{e_1e_2,i_1i_2}h_{i_1i_2}$
with $\grp(\bm{t})=\langle (12)(34)\rangle$ and
$\grp(\bm{h})=\langle (12)\rangle$, we have $S=(\bm{t},\bm{h})$,
$\grp(S)=\langle (12)(34),(56)\rangle$,
$\pi=\{e_1,e_2,i_1,i_2,i_1,i_2\}$,
$\varpi(\pi)=\{\theta(3,5),\theta(4,6)\}$. Then
$\grp_{\Omega_I(\pi)}(S)=\grp_{\Omega_E(\pi)}(S)=\grp(S)$ and
$\grp_{\varpi(\pi)}(S)=\langle (12)(34)(56)\rangle$ such that
$\grp(\Omega_E(\pi))=\langle(12)\rangle$ and $\Ker\phi=\langle
e\rangle$. Obviously, th permutation $(12)\notin\grp(S)$, i.e.,
$\grp(\Omega_E(\pi))$ is not a subgroup of $\grp(S)$. Only through
the homomorphism $\phi$ \eqref{Homo}, we can get the permutation
symmetry group of $Z^E$.

\section{Graphical canonicalization algorithms}\label{algorithm}
\subsection{Traditional backtrack algorithm}\label{traditionalBacktrack}
As one can see from the above examples, to compute the canonical
forms of tensor products \eqref{tensorProduct}, it is essential to be
able to calculate the group chain such as $\grp_{\varpi(\pi)}(S)$, which can determine all
possible contraction patterns via the left coset decomposition
\eqref{lcr}. The structure of $\varpi(\pi)$ reveals it as a special
combinatorial object. For this kind of problems, the backtrack
searching\cite{kreher1998combinatorial,seress2003permutation} appears to be the only possible approach, which potentially involves searching through all of
group elements of a permutation group, and hence the computational
complexity is at least $O(|\grp(S)|)$ in the worst case.
To make it work in practice, in designing such
algorithms it is crucial to find methods to skip as many group elements as
possible during the search. This is often referred as \emph{pruning the
search tree}. The backtrack algorithms have also been used in
problems including centralizers and normalizers of elements and
subgroups, stabilizers of subsets of $\Omega$, and intersections of
subgroups. For more applications, we refer the readers to Refs.
\cite{seress2003permutation,holt2005handbook}.

\begin{figure}\centering
\resizebox{0.45\textwidth}{!}{\includegraphics{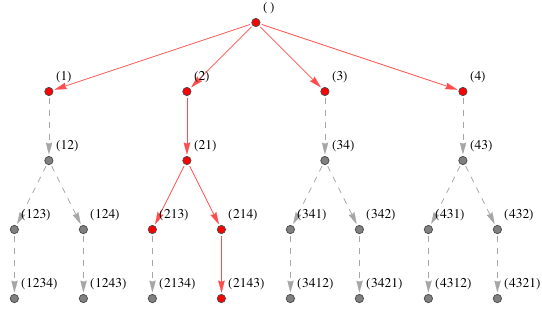}}   \\
\caption{Search tree of the modified traditional backtrack in
finding representative for $g_{e_4e_1,e_3e_2}$ with
$\grp(\bm{g})=\langle (12),(34),(13)(24)\rangle$. The red parts
represent the visited nodes and the actual search path, while all
the gray parts are pruned. The final result is
$\mathcal{C}(g_{e_4e_1,e_3e_2})=g_{e_1e_4,e_2e_3}$.}\label{searchtree}
\end{figure}

Before introducing our algorithm for general cases,
we first show that in the special case (S\ref{S2}) with no internal indices, the canonical form can be efficiently in polynomial scaling
by a simple modification of the traditional backtrack. Given
$\Omega=\{1,2,\cdots,D\}$ and $\grp=\grp(S)$, denoting the pointwise
stabilizer of the $i-1$ first elements of $\Omega$ (represented by $\Omega_{i-1}$)
by $\grp^{[i]}\triangleq \grp_{(\Omega_{i-1})}$, we have
a \emph{stabilizer chain}, viz.,
\begin{eqnarray}
\grp=\grp^{[1]}\geq \grp^{[2]}\geq\cdots
\geq\grp^{[D]}\geq\grp^{[D+1]}=\langle e\rangle.
\end{eqnarray}
Let $U^{(k)}$ be a left transversal (a set of left coset
representatives) for $\grp^{[k+1]}$ in $\grp^{[k]}$. Then every
element $g\in\grp$ can be uniquely decomposed into
\begin{eqnarray}
g=u_{1,i_1}u_{2,i_2}\cdots u_{D,i_D},\quad u_{k,i_k}\in
U^{(k)},\label{SSrep}
\end{eqnarray}
and $|\grp|=\prod_{k=1}^{D}|U^{(k)}|$. This is
referred as the Schreier-Sims representation\cite{kreher1998combinatorial} of
$\grp$. Based on Eq. \eqref{SSrep},
a backtracking procedure (depth-first transversal) can be used to run
through the elements of $\grp$, which results in an organization of
the elements of $\grp$ as a search tree $\mathcal{T}$,
see Figure \ref{searchtree} for an example.
More specifically, the root at level 0 labeled by
empty represents $\grp=\grp^{[1]}$, The nodes at level $k=1$
represent the coset $u_{1,i_1}\grp^{[2]}$ ($i_1=1,\cdots,|U^{(1)}|$)
labeled by $(\gamma_{i_1})$, where $\gamma_{i_1}=u_{1,i_1}(1)$
is the image of 1 under the action of the permutation $u_{1,i_1}$.
In general, every node at level $k>0$ is labeled with a sequence
$(\gamma_{i_1},\cdots,\gamma_{i_k})\subseteq \Omega$ referred as \emph{partial images}, which represents the coset $u_{1,i_1}\cdots u_{k,i_k}\grp^{[k+1]}$.
The node $(\gamma_{i_1},\cdots,\gamma_{i_k})$ has $|U^{(k+1)}|$ children
$(\gamma_{i_1},\cdots,\gamma_{i_k},\gamma_{i_{k+1}})$ for each
$\gamma_{i_{k+1}}\in (\Delta^{(k+1)})^g$,
where $(\Delta^{(k+1)})^g$ denotes the image of
the set $\Delta^{(k+1)}$ under the action of $g$,
$\Delta^{(k+1)}$ is the orbit
of $k+1$ under the action of $\grp^{[k+1]}$,
and $g$ is an arbitrary permutation fulfilling
$g(1,\cdots,k)=(\gamma_{i_1},\cdots,\gamma_{i_k})$.
Therefore, at the level $D$, the leaves correspond to all the elements of $\grp$.
Each path from the root to a leaf represents a sequence of group
elements $u_{1,i_1}$, $(u_{1,i_1}u_{2,i_2})$,
$(u_{1,i_1}u_{2,i_2}u_{3,i_3})$, \ldots, $(u_{1,i_1}u_{2,i_2}\cdots
u_{D,i_D})$, and from one element to the next element, one more
image of point in $\Omega$ is fixed.

Now suppose the coloring
$\pi(\Omega)=\{\pi(1),\pi(2),\cdots,\pi(D)\}$ with all $\pi(i)$ are
different, then $\grp_{(\pi)}(\Omega)=\langle e\rangle$ and
$\grp(\Omega(\pi))=\grp(S)$. Especially, the elements in
$\Class{\pi}=\{g\circ\pi(\Omega):g\in\grp(S)\}$ have a one-to-one
correspondence with the group elements $g\in\grp(S)$. Then, the
lexicographical order of colors $\pi(i)$ leads to a natural
ordering of the partial images like
$(\gamma_{i_1},\cdots,\gamma_{i_k})$ at the same level $k$. That is,
we say
$\gamma_k=(\gamma_{i_1},\cdots,\gamma_{i_k})<\gamma_k'=(\gamma_{i_1'},\cdots,\gamma_{i_k'})$
if $\pi(\gamma_k)<\pi(\gamma_k')$. In particular, there is only one
minimum $\pi$ that satisfies (C\ref{c3}), i.e., only one minimum
$(\gamma_{i_1},\cdots,\gamma_{i_D})$ at the level $D$ of the search
tree $\mathcal{T}$. Most importantly, there is also a \emph{unique} minimum
at each level $k$ of $\mathcal{T}$, which is simply obtained from
the first $k$ elements of the minimum at level $k+1$. Therefore, we
can prune the search tree by only retaining the \emph{minimal partial image}
at each level. This suggests a modification of the traditional
backtrack search, which is a uniform depth-first search, into a
guided depth-first search. The order of the nodes to be visited in
the next level is obtained by first taking a local breadth-first search
from the current node, and then comparing the partial images. This
is illustrated in Figure \ref{searchtree} for a simple example $g_{e_4e_1,e_3e_2}$
with $\grp(\bm{g})=\langle (12),(34),(13)(24)\rangle$. The final
result is given by $\pi_{\canon}=\pi(\{2,1,4,3\})=\{e_1,e_4,e_2,e_3\}$, i.e.,
$\mathcal{C}(g_{e_4e_1,e_3e_2})=g_{e_1e_4,e_2e_3}$. Note that a
large part of the search tree has been pruned, see
the gray nodes in Figure \ref{searchtree}.

In fact, for $\grp(S)=\mathcal{S}_D$, the symmetric group of degree $D$, this
procedure is similar to the selection sort for $\pi$, which has
$O(D^2)$ computational complexity instead of factorial
$O(|\grp(S)|)=O(D!)$. In general, since at
the level $k$ the number of point images that have been fixed are $k$, and the
number of children for a node at level $k$ will not exceed $D-k$,
we can conclude that the number of nodes that will be visited
by this algorithm for the external indices only case
is at most $O(D^2)$.

\subsection{Graphical representation of tensor products}
Unlike for the special case (S\ref{S2}), in which there is always a unique
minimum at each level, in the presence of internal indices, several
nodes at the same level may need to be explored, when $\Ker\phi\ne\langle
e\rangle$. Moreover, in this case it is not natural to define an
ordering for partial images, which is of significant importance in
pruning the search tree. This creates difficulties in using the
traditional back track algorithm for efficiently finding
the canonical form.

The pair structure of elements in $\varpi(\pi)$ suggests that it is better to be viewed as a list of edges in a graph. Moreover, if there are external indices in $\pi$,
then we can enlarge $\Omega$ to contain $D+1$ natural numbers. The first
element 1 is chosen to correspond to an auxiliary vertex,
while the rest $D$ elements correspond to those in the original $\Omega$.
[NB: This setting will ensure that a representative consistent with the conditions in Theorem \ref{representative}, in particular, the condition (C\ref{c1}), will be found.]
Consequently, we can augment $\varpi(\pi)$ with $\theta(1,i)$ for $i\in\Omega_E(\pi)$,
then the problem of finding the canonical form satisfying
(C\ref{c1})-(C\ref{c4}) can be solved in a single framework.
This will be similar to the canonicalization of graphs\cite{mckay2014practical,piperno2008search},
but with some important differences that will be mentioned later.
Before introducing the algorithm, we formalize the
correspondence between tensor products and graphs more explicitly.

\begin{dfn}[graphical representation of tensor products]
An undirected graph is an ordered pair $G=(\Vset,\Eset)$, where
$\Vset$ is a finite set of vertices or nodes, and $\Eset$ is a set
of unordered pairs of vertices called edges. For a tensor product of
form \eqref{tensorProduct}, suppose its correspondent coloring is
$\pi$, we associate $\pi$ with a graph $G$ resulting a colored/labeled graph composed of $\Vset=\Omega$
and $\Eset=\{(1,j):j\in\Omega_E(\pi)\}\cup \{(i,j):i,j\in
\Omega_I(\pi)\}$ for general cases. The graph in special cases
(S\ref{S1}) and (S\ref{S2}) can be obtained similarly.
\end{dfn}

\begin{figure}\centering
\begin{tabular}{cc}
\resizebox{0.22\textwidth}{!}{\includegraphics{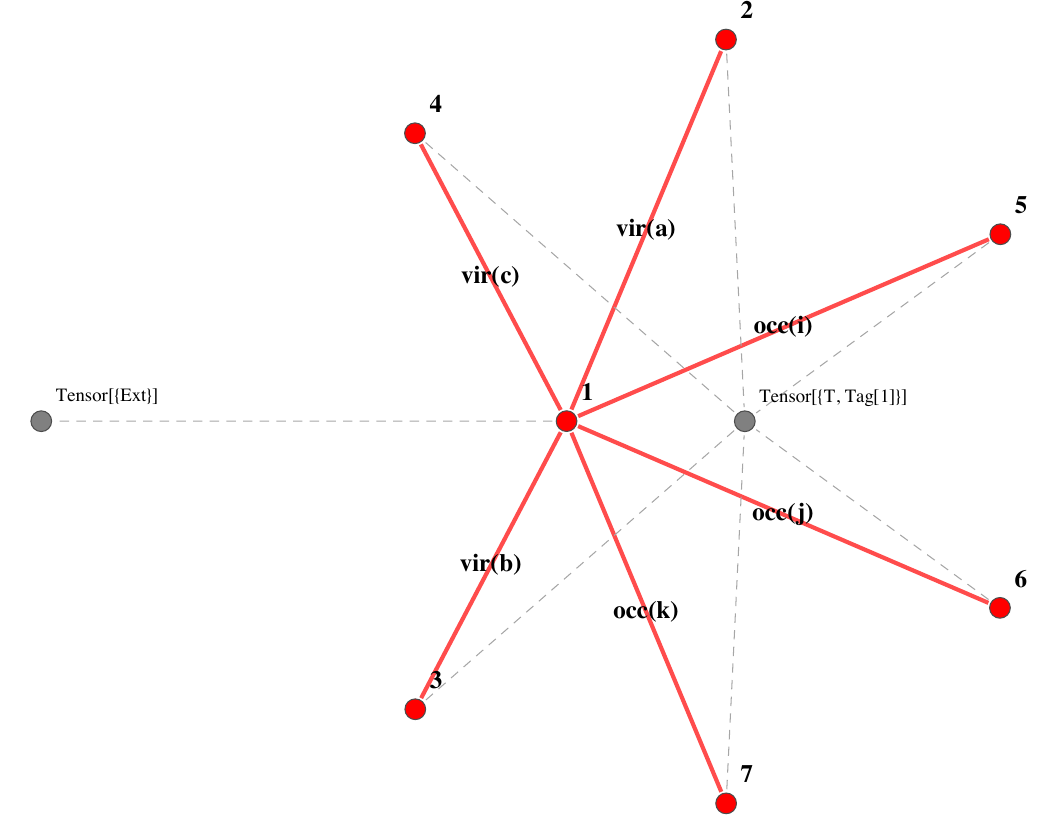}} &
\resizebox{0.22\textwidth}{!}{\includegraphics{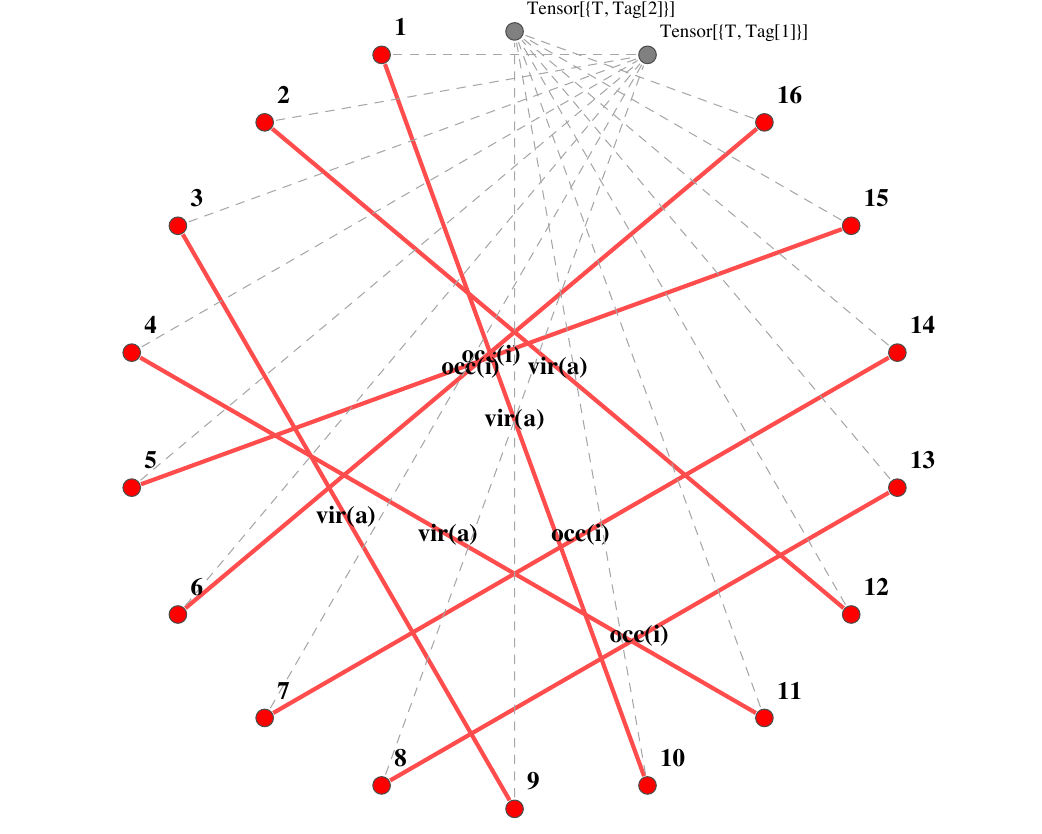}} \\
(a) $T^{abc}_{ijk}$  & (b) $T^{abcd}_{ijkl}T^{cadb}_{lkij}$  \\
\resizebox{0.22\textwidth}{!}{\includegraphics{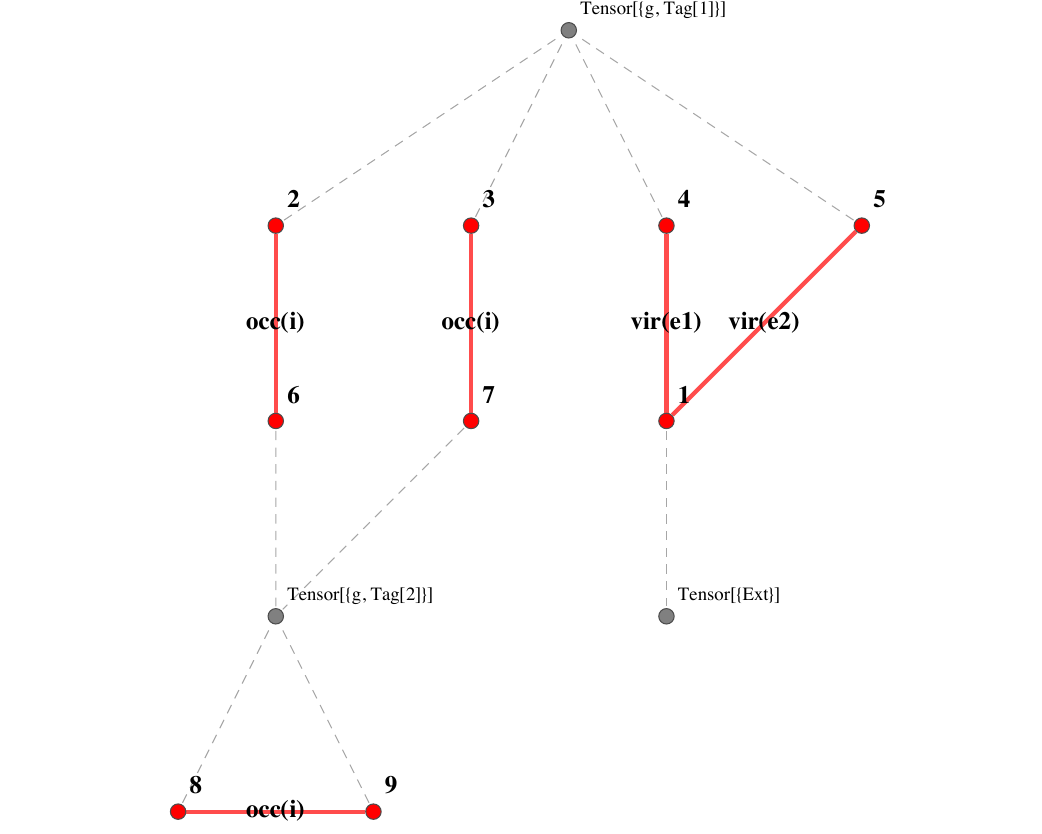}} &
\resizebox{0.22\textwidth}{!}{\includegraphics{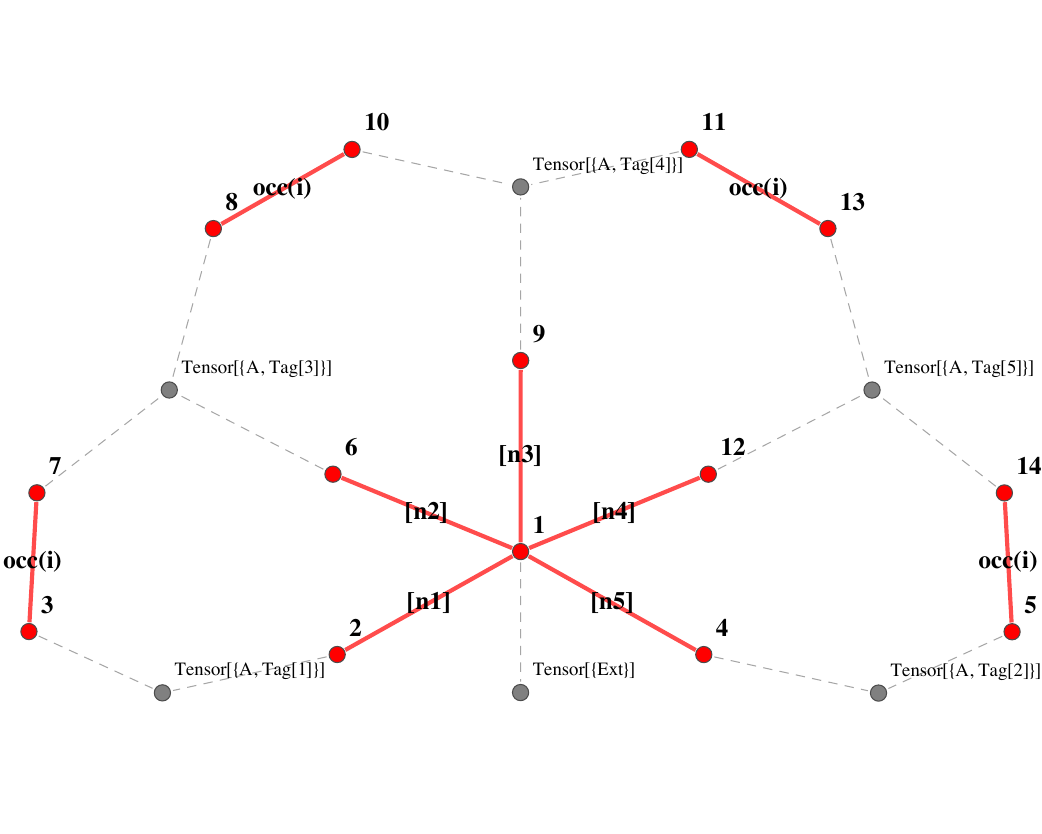}} \\
(c) $g_{i_1i_2,i_3i_3}g_{i_1i_2,e_1e_2}$ &
(d) $A^{n_1}_{i_1}A^{n_2}_{i_1i_2}A^{n_3}_{i_2i_3}A^{n_4}_{i_3i_4}A^{n_5}_{i_4}$ \\
\resizebox{0.18\textwidth}{!}{\includegraphics{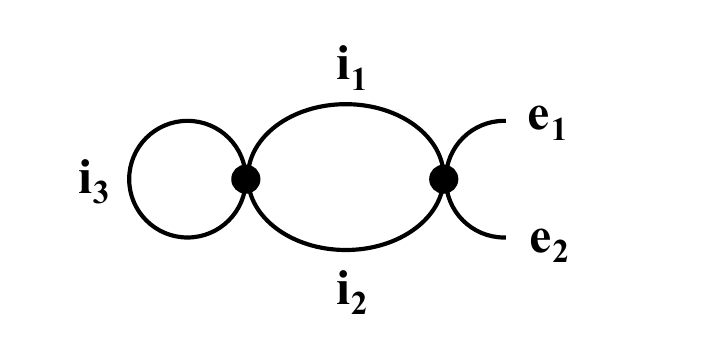}} &
\resizebox{0.18\textwidth}{!}{\includegraphics{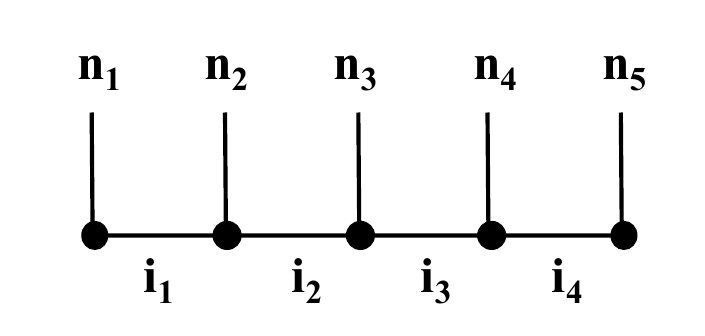}} \\
(e) Hugenholtz diagram & (f) MPS diagram \\
\end{tabular}
\caption{Graphical representation of tensor expressions (a)-(d). The red parts
form the graphs $G=(\Vset,\Eset)$, while the gray parts are not part of $G$
and just displayed to reveal the origin of vertices $\Vset$ from the tensors.
The labels for $\Eset$ are shown explicitly.
In the case of internal indices, only their types are shown as the labels.
The diagrams in perturbation theory (e) and tensor network states (f)
are also depicted in comparison with the same terms (c) and (d), respectively.}\label{graphrep}
\end{figure}

In Figure \ref{graphrep}, we show some examples for the introduced graphical representation of tensor products. These include (a) a
triple excitation operator $T^{abc}_{ijk}$,
(b) a fully contracted term
$T^{abcd}_{ijkl}T^{cadb}_{lkij}$,
(c) the example studied in Sec. \ref{example},
$g_{i_1i_2,i_3i_3}g_{i_1i_2,e_1e_2}$,
and (d) a 5-site matrix product state (MPS) as a special case of tensor network states\cite{chan2012low,orus2014practical}.
This graphical representation
is inspired by diagrammatic methods in permutation theory\cite{paldus1975}
and tensor network diagrams\cite{chan2012low,orus2014practical},
for comparison, see the conventional Hugenholtz diagram (e) and
MPS diagram (f) representing the same quantities as (c) and (d), respectively.
It is clear that these graphs (a)-(d) have extremely simple structures, namely except for the auxiliary vertex whose
degree can be larger than one, the degree of other vertices is just
one. In addition, there are many disconnected components, if the number of
internal indices is large, see Figure \ref{graphrep}(b).
We can label the vertex set $\Vset$ in
accord with the factor set $S$, while there are some freedoms to label
the edge set $\Eset$. The labels for $\Eset$ can be chosen as the
information that fully specifies the index pair $(i,j)$. In this
work, we label the pair by a pattern '$type[index]$', where
the '$type$' specifies the type of $i$ or equivalently $j$ (e.g., '$occ$' for occupied
orbitals and '$vir$' for virtual orbitals used in Figure \ref{graphrep}) and '$index$' can be the specific index (e.g., '$e_1$' and '$i_1$') appeared in the expressions or general unspecific indices (e.g., '$i$' for occupied and '$a$' for virtual).
In this convention, we have the label
$\theta(i,j)=\{(i,j),type[index]\}$ for an edge $(i,j)$.
[NB: This choice will prioritize $(i,j)$ over $type[index]$ in the comparison
of two edge sets.] Depending on whether the '$index$' takes the value of specific or general index, we call the corresponding graph \emph{labeled} or \emph{unlabeled},
respectively.

\begin{dfn}[labeled and unlabeled graphs]
For a given tensor product \eqref{tensorProduct}, or equivalently, a
factor set $S$ and a coloring $\pi$, we have the following four kinds of graphs:
\begin{enumerate}[(K1)]
\item\label{k1} The externally labeled, internally labeled graphs $G^{EI}(\pi)$ with labeled edge set $\Eset^{EI}(\pi)$, which have a one-to-one correspondence
    with the tensor product expressions.

\item\label{k2} The externally labeled, internally unlabeled graphs $G^{E}(\pi)$
with labeled edge set $\Eset^{E}(\pi)$, which will be termed as diagrams,
see Figure \ref{graphrep}(c).

\item\label{k3} The externally unlabeled, internally labeled graphs
$G^{I}(\pi)$ with labeled edge set $\Eset^{I}(\pi)$.

\item\label{k4} The externally unlabeled, internally unlabeled graphs $G(\pi)$ with unlabeled edge set $\Eset(\pi)$, which will be termed as skeletons.
\end{enumerate}
\end{dfn}

In accord with these labeling schemes, we define the corresponding
label-preserved graph isomorphism as follows
\begin{dfn}[label-preserved graph isomorphism]\label{labelPreservedGI}
Two labeled graphs $G_1=(\Vset_1,\Eset_1)$ and
$G_2=(\Vset_2,\Eset_2)$ are identical, i.e., $G_1=G_2$, if
$\Vset_1=\Vset_2$ and $\Eset_1=\Eset_2$ in the sense that
$\theta(u,v)\in \Eset_1$ if and only if $\theta(u,v)\in \Eset_2$.
Two graphs $G_1=(\Vset_1,\Eset_1)$ and $G_2=(\Vset_2,\Eset_2)$ are
isomorphic under $\grp(S)$, i.e., $G_1\cong G_2$, if $\exists
g\in\grp(S)$ such that $\theta(g(x),g(y))\in \Eset_2$ if and only if
$\theta(x,y)\in \Eset_1$, i.e., $g\circ \Eset_1=\Eset_2$.
\end{dfn}
Consequently, the automorphism of the labeled graph is defined as
\begin{dfn}[automorphism of graph]\label{autoLG}
If $g\in\grp(S)$, such that $g\circ G\triangleq
(\Vset,g\circ\Eset)=G$, i.e., $g\circ\Eset=\Eset$, then $g$ is an
isomorphism from a graph $G$ to itself, which is called an
automorphism. The set of all automorphisms of $G$ in $\grp(S)$ forms
the automorphism group $\Aut_{\grp(S)}(G)$ of the labeled graph. We
use the subscript to emphasize that the automorphism group is
calculated from $\grp(S)$. The action of $\Aut_{\grp(S)}(G)$ on
$\Vset$ partitions $\Vset$ into orbits, and induces an equivalence
relation on $\Vset$: two vertices $u$ and $v$ are equivalent if and only if they
are in the same orbit, i.e., there exists an automorphism
$g\in\Aut_{\grp(S)}(G)$ such that the image of $u$ under $g$ is $v$,
i.e., $u^g=v$.
\end{dfn}
The canonical form of graph is defined as
\begin{dfn}[canonical form of graph]\label{canonicalGraph}
The canonicalization is a mapping such that for all $g\in \grp(S)$
and graph $G$,
(1) $\mathcal{C}(G)\cong G$ and (2)
$\mathcal{C}(G^g)=\mathcal{C}(G)$. By the property (2), which is
called ''label-invariance'', the image $\mathcal{C}(G)$, called a
canonical form, is a unique representative of its isomorphic class
$\{g\circ G:g\in\grp(S)\}$. The importance of canonical form is that
$G\cong G'$ if and only if $\mathcal{C}(G)=\mathcal{C}(G')$.
\end{dfn}
For our purpose, the natural definition is that the
canonical graph has the smallest \emph{sorted} labeled edge set
$\Eset(\pi)$, where the sorting can be simply based on the lexicographical
order of $\theta(i,j)$. Clearly, such definition is the counterpart
of that defined in Theorem \ref{representative}. From these
definitions, we can have the following important theorem.

\begin{thm}\label{connectionFinal}
The connections between tensor product canonicalization and
graph canonicalization defined by Definitions \ref{labelPreservedGI}-\ref{canonicalGraph} are as follows:
\begin{enumerate}
\item
Two tensor products are symmetry equivalent (or two colorings $(S_1,\pi_1)$
and $(S_2,\pi_2)$ are $(\grp(S),\subgrp)$-equivalent), if and only
if for diagrams $G^E(\pi_1)\cong G^E(\pi_2)$.

\item The computation of a representative
for $\pi$ defined in Theorem \ref{representative} is equivalent to
the computation of canonical form for $G^E(\pi)$
and relabelling internal indices for $G^E(\pi)$ afterwards.

\item In particular, we
have $\grp_{\varpi(\pi)}(S)=\Aut_{\grp(S)}(G(\pi))$
for the skeleton $G(\pi)$ and
$\Ker\phi=\Aut_{\grp(S)}(G^E(\pi))$ for the diagram $G^E(\pi)$,
which follows from the fact that $\Ker\phi$ is the pointwise stabilizer of $\Omega_E(\pi)$ in $\grp_{\varpi(\pi)}(S)$.
\end{enumerate}
\end{thm}

The importance of this theorem is that to compute the representative
of tensor products, we can simply compute the canonical form of the
corresponding externally labeled graph $G^E(\pi)$ and then perform
a relabeling of the internal indices. If the
permutation symmetry group is also required, we can calculate it from the
quotient group $\grp_{\varpi(\pi)}(S)/\Ker\phi=
\Aut_{\grp(S)}(G(\pi))/\Aut_{\grp(S)}(G^E(\pi))$ via
the homomorphism $\phi$ \eqref{Homo},
where both $\Aut_{\grp(S)}(G(\pi))$ and $\Aut_{\grp(S)}(G^E(\pi))$
can be computed from the same algorithm (vide post)
by inputting $G(\pi)$ and $G^E(\pi)$, respectively.

In sum, we have reformulated the tensor canonicalization problem
into a graph canonicalization problem. However, we should emphasize
that the graph isomorphism defined in Definition \ref{labelPreservedGI}
is different from the standard graph isomorphism problem,
in which an arbitrary permutation of the vertices
is allowed, viz., $\grp(S)=\mathcal{S}_D$.
This is a fundamental difference. Because
in view of the simplicity of graphs in
Figure \ref{graphrep}, the isomorphism of two graphs sharing the
same kind of factor set can be simply checked
by the number of internal and external indices.
However, the problem with a restricted set of permutations given by
$\grp(S)$ is obviously more difficult.
However, this graphical reformulation does have advantages over the formulation by double coset representatives\cite{portugal1999algorithmic,portugal2001group,
manssur107032group,manssur2002group,manssur2004canon,martin2008xperm,niehoff2018faster}.
Because the latter is known to be exponential\cite{butler1991fundamental},
while it may be possible to develop polynomial scaling algorithm
for the introduced graphs.
In fact, for graphs with bounded degree, the standard
graph isomorphism problem is of polynomial complexity\cite{luks1982isomorphism}.
Although the permutations are limited in our case, it is reasonable to be optimistic
to solve the graph canonicalization problem and hence
the tensor canonicalization problem in polynomial scaling.
In fact, we will show that for the following algorithm
we introduced is \emph{polynomial}
for the worst case in the Butler-Portugal algorithm.

\subsection{Partition backtrack algorithm}\label{sec:ALGORITHM}
Our algorithm for computing the canonical
form of a graph and generators of its automorphism group
is based on the partition backtrack algorithm used
in state-of-the-art packages for graph isomorphism problem via computing canonical labelings\cite{mckay2014practical,piperno2008search}.
However, the difference in symmetry group results in some notable
changes in some parts of the partition backtrack algorithm,
in particular, in defining proper refinement procedure, see Sec. \ref{sec:refinement},
which must ensure that the used permutation is indeed in $\grp(S)$.
To take into such difference, our method for exhausting the elements of the group is similar to that used in traditional backtrack. But instead of working with partial images, we will work with partitions of $\Vset$ similar to the algorithm for graphical isomorphism.
To enhance the pruning of search tree, the local breadth-first search introduced in
Sec. \ref{traditionalBacktrack} for the modified traditional backtrack is also used here.
In case of large $\Aut_{\grp(S)}(G^E(\pi))$, which potentially enlarges the
branching factors of  the search tree, the idea of using
automorphism to prune the search tree developed for the general
isomorphism problem\cite{mckay2014practical,piperno2008search} is employed.

\subsubsection{Search tree based on partitions}
Most of the graph isomorphism algorithms employed the same
individualization-refinement paradigm but differ in some
details. The central quantity is the partition.
\begin{dfn}[partition]
An ordered partition of the set $\Vset$ is a sequence of subsets
$\Pi=(\Pi_1,\Pi_2,\cdots,\Pi_r)$, such that
$\Pi_i\ne\emptyset$,$\Vset=\bigcup_{i=1}^r \Pi_i$, and $\Pi_i\cap
\Pi_j=\emptyset$ for $i\ne j$. The ordered sets $\Pi_i$ are called cells of
$\Pi$. A discrete ordered partition is an ordered partition with
each cell being a singleton $|\Pi_i|=1$.
\end{dfn}
The following relation defines a partial order for the set of all
ordered partitions $\Pi(\Vset)$.
\begin{dfn}
For $\Pi_1,\Pi_2\in\Pi(\Vset)$, we say $\Pi_2$ is finer than
$\Pi_1$, denoted by $\Pi_2\preceq\Pi_1$, if each cell of $\Pi_1$ is
a consecutive union of cells of $\Pi_2$.
\end{dfn}
The search tree can be constructed by the individualization and refinement procedures.
\begin{dfn}[individualization]
Let $v\in\Vset$ belong to a non-singleton cell $\Pi_i$ of an ordered
partition $\Pi$, then $\Pi\downarrow
v=(\Pi_1,\cdots,\Pi_{i-1},\{v\},\Pi_{i}\backslash\{v\},\Pi_{i+1},\cdots,\Pi_r)$
denotes the partition obtained from $\Pi$ by splitting $\Pi_i$ into
the cells $\{v\}$ and the complement $\Pi_{i}\backslash\{v\}$. We
call $\Pi\downarrow v$ is obtained from $\Pi$ by individualizing
vertex $v$. Obviously, the relation $(\Pi\downarrow v)\prec\Pi$ holds.
\end{dfn}

Before discussing the refinement procedure, we describe how to construct
the search tree in a partition backtrack algorithm.
Note that the partition $\Pi$ can be viewed as a coloring for
$\Vset$ (not to be confused with $\pi$). A vertex colored graph is a
pair $(G,\Pi)$, where $G$ is a graph and $\Pi$ is a coloring. For an
initially-specified colored graph $(G,\Pi_0)$, the search tree
$\mathcal{T}(G,\Pi_0)$ is constructed by selecting the first
non-singleton cell of $\Pi_0$, individualizing it in all possible
ways allowed by the symmetry group,
refining the new partition to new nodes $R(G,\Pi,v)$, and
finally terminating when the leaves, i.e., the discrete partition,
are reached.  Note that the leaves have a one-to-one
correspondence with the elements of $\grp(S)$, because the discrete
partitions are just images of $\Vset$ under the action of $\grp(S)$.
This is similar to the traditional backtrack discussed before.
Thus, after traversing the whole search tree, the canonical graph
defined by having the \emph{smallest} sorted labeled edge set $\Eset(\pi)$
in accord with Theorem \ref{representative} can be found.
Besides, the automorphisms can also be found at the leaves, because for two
discrete partitions $\Pi_1=g_1(\Vset)$ and $\Pi_2=g_2(\Vset)$ with
their correspondent $\Eset(\pi)$ being identical, then the permutation
$g_1^{-1}g_2$ is an automorphism.

By implementing the above procedure, the canonical form and
automorphism group can be computed with a full search tree,
whose size can be exponential for large $\grp(S)$.
To make this procedure practical, following the ideas
in graph isomorphism algorithms\cite{mckay2014practical,piperno2008search},
we could use the refinement procedure and pruning techniques
based on the non-discrete partitions to reduce the size of search tree.

\subsubsection{Refinement}\label{sec:refinement}

\begin{dfn}[refinement]
A refinement of $(G,\Pi)$ is a partition $\mathcal{R}(G,\Pi)$ such
that (i) $\mathcal{R}(G,\Pi)=(G,\Pi')$ where $\Pi'\preceq\Pi$, (ii)
$\mathcal{R}$ preserves isomorphisms, which means if
$(G_1,\Pi_1)\cong(G_2,\Pi_2)$, then $\mathcal{R}(G_1,\Pi_1)\cong
\mathcal{R}(G_2,\Pi_2)$.
\end{dfn}
Due to the simple structure of our graphs and restrictions on permutations
by $\grp(S)$, we proposed an refinement procedure as follows:

\begin{dfn}
Given $(G,\Pi)$ and the associated subgroup
$\grp_{\Pi}(S)=\{g\in\grp(S): \Pi^g_i=\Pi_i,\Pi_i\in\Pi\}$, the
refinement $\mathcal{R}(G,\Pi)$ is obtained by a repeated
application of the following two operations until the partition is
not changed:
\begin{enumerate}[(R1)]
\item \label{r1} If the first vertex $v$ in a non-singleton cell is stabilized by
$\grp_{\Pi}(S)$, then it can be singled out which leads to a new
partition $\Pi\downarrow v$.

\item \label{r2} Suppose $(\{v_1\},\{v_2\},\cdots,\{v_k\})$ are the first
$k$ singleton cells of $\Pi$, such that $\Pi_{k+1}$ is a
non-singleton cell, then we consider these singleton cells
sequentially. Suppose $v_i$ is being visited, then
its neighbor in the graph $G$ is examined:
\begin{enumerate}[(a)]
\item If its neighbor contains more than one
element (in our graph this can only be the case for $v_1=1$ when
there are external indices), then the nonsingleton
cells wherever the neighbor element lies in are marked such
that they will not be modified in the refinement procedure.

\item If its neighbor is a single vertex $u_i$ and $u_i$ is also in a
singleton cell in $\Pi$, then we move to consider the next vertex $v_{i+1}$.

\item If its neighbor is a single vertex $u_i$, which is in a non-singleton
cell and can be moved to the first element by a permutation $g\in\grp_{\Pi}(S)$,
then we obtain a new refined partition $\Pi^g\downarrow u_i$, otherwise, $\Pi$ is returned. [NB: In the case that the non-singleton cell corresponds
to several identical factors $\bm{X}$, a coarser split of
the non-singleton cell is first applied to individualize the subcell
where $u_i$ is in.]
\end{enumerate}
\end{enumerate}
\end{dfn}
Using this refinement procedure, usually the depth of search tree is reduced
without affecting the computation of canonical form and automorphism group.
As an example, for the expression with only internal indices, without refinement
the depth of the search tree is at most $D$, while with refinement
the depth is at most $D/2$. The refinement procedure used here
is by no means optimal, but it is sufficient for our examples
illustrated below.

\subsubsection{Pruning with non-discrete partition}
\begin{dfn}[position]
The position of a vertex $v\in\Vset$ in an ordered partition $\Pi$
is defined by $p(v,\Pi)=1+\sum_{i=1}^{k-1}|\Pi_i|$ for $v\in\Pi_k$.
\end{dfn}
With this definition and $\Eset(\pi)$, we can introduce a function
of $\Pi$,
\begin{eqnarray}
\Eset(\pi,\Pi)=\{\theta(p(i,\Pi),p(j,\Pi)):i,j\in\Vset,\;(i,j)\in\Eset(\pi)\}.\label{edgeset}
\end{eqnarray}
This function is actually the edge set of the quotient graph
$Q(G,\Pi)=\{\Vset',\Eset'\}$, with $\Vset'=\{p(v,\Pi):v\in\Vset\}$
and $\Eset'=\{\theta(p(u,\Pi),p(v,\Pi)):
\theta(u,v)\in\Eset(\pi)\}$, if we consider the partition $\Pi$ as
an equivalence relation on $\Vset$, namely, $u,v\in\Vset$ are called
equivalent if they are in the same cell, i.e., $u,v\in\Pi_i$ for
some $i$. The so-defined function $\Eset(\pi,\Pi)$ has an important
property,
\begin{thm}\label{ePathToLeaf}
If $\Pi_1\prec\Pi_2$, then $\Eset(\pi,\Pi_1)>\Eset(\pi,\Pi_2)$.
\end{thm}
Its correctness can be verified by considering a simple example.
This shows that along a path from the root to a leaf, the value of
$\Eset(\pi,\Pi)$ is increasing. That is, the value of
$\Eset(\pi,\Pi)$ at a given node $\nu$ is a \emph{lower bound} for all the
values $\Eset(\pi,\Pi)$ of its descendent. Therefore, if at the node
$\nu$, its corresponding $\Eset(\pi,\Pi)$ is larger than the minimal
$\Eset_{\min}(\pi)$ we currently have, then the entire subtree headed
at $\nu$ can be pruned. To make this pruning more
powerful, a local breadth-first search is applied at the current node
$\nu$ to reorder the children to be visited in the following
depth-first search in an increasing order of $\Eset(\pi,\Pi)$.

\subsubsection{Pruning with automorphism}
The pruning based on automorphism group is crucial for graphs having
large automorphism groups. The basic idea is simple. At a give node
$\nu$ in $\mathcal{T}(G,\Pi_0)$, suppose we have an subgroup
$\Gamma$ of $\Aut_{\grp(S)}(G(\pi))$ at hand, then by applying its
elements to the partition $\Pi$, we obtain several new partitions
$\Pi^g$. If some $\Pi^g$ has been visited before, then the entire
subtree $\mathcal{T}(G,\Pi_0,\nu)$ is the same as that visited
subtree, such that it can be pruned. This pruning ensures that only
the \emph{generators} of the automorphism group will be found, rather than all
the group elements during the backtrack searching, which can be enormous
for tensor products with many internal indices.
However, the computational cost of a naive implementation based on checking the
action of every element of the subgroup will still scale as $O(n!)$
when the order of the automorphism group scales as $O(n!)$ for large
$n$. This is the case for the kind of tensor expressions in Figure
\ref{graphrep}(b). Therefore, although the number of visited
intermediate nodes are small, the time for checking can be very
long.

The solution to this problem is also based on a classification of elements in
$\Gamma$. Suppose the parent of the node $(\nu,\Pi)$ is
$(\nu',\Pi')$ and the stabilizer of the ordered partition $\Pi'$ is
$\Gamma_{\Pi'}=\{g\in\Gamma:g(\Pi'_i)=\Pi'_i,\Pi'_i\in\Pi'\}$,
we can have the coset decomposition $\Gamma=\bigcup_r g_r
\Gamma_{\Pi'}$. The meaning of this decomposition is clear: for
$g\in\Gamma_{\Pi'}$, it only transforms the branches of the subtree
$\mathcal{T}(G,\Pi_0,\nu')$, while the left coset representative
transform the entire subtree to another subtree. Now suppose
$\exists g\in\Gamma$ such that $\Pi^g=g\circ\Pi$ has been visited
before. Let the deepest common ancestor of $\Pi$ and $\Pi^g$ in the
search tree $\mathcal{T}(G,\Pi_0)$ be denoted by $\Pi_*$, and the
subtrees contain $\Pi$ and $\Pi^g$ be $\mathcal{T}$ and
$\mathcal{T}^g$, respectively, there can only be two cases: (1)
$g\notin\Gamma_{\Pi'}$ or (2) $g\in\Gamma_{\Pi'}$. These two cases
lead to two different kinds of pruning based on the detected
automorphisms developed in graph canonicalization algorithms\cite{mckay2014practical}:
\begin{enumerate}[(P1)]
\item\label{p1} In the first case, $g$ must have be found during the search of
$\mathcal{T}$ before visiting $\Pi$. Actually, once such $g$ is
found at a leaf node of $\mathcal{T}$, we can trace back to $\Pi_*$
and prune the entire subtree $\mathcal{T}$.

\item\label{p2} If such pruning has been employed, then at a given node, only
the second case is left, in which instead of the full automorphism
group $\Gamma$, only the subgroup $\Gamma_{\Pi'}$ needs to be
considered. Moreover, in this case, we only need to examine
whether there is an element in the orbit of $\Pi$ under the action
of $\Gamma_{\Pi'}$ that has been visited before.
\end{enumerate}
In sum, by taking these two economic pruning strategies, only the
paths that lead to leaves corresponding to generators of the
automorphism group are retained, while all the parts corresponding
to a composition of generators can be pruned. Thus, even in the presence of a
large automorphism group, in which case the pruning based on
$\Eset(\pi,\Pi)$ takes no effect, the search space can still be
significantly reduced.

\subsubsection{Algorithm and possible improvements}\label{sec:improvement}
The pseudocode of our final algorithm with pruning is
presented in Algorithm \ref{alg}. A preliminary implementation of this algorithm
has been made into a package named \textsc{CanonicalTensorProducts} using \textit{Mathematica}\cite{Mathematica}. Our implementation
is proof-of-principle, and many possible improvements
can be applied. For instance, the shape of search tree depends crucially
on the order of tensors and definitions of canonical form.
Besides, better refinement functions may be designed.
Other searching strategies used in the graph isomorphism
algorithms\cite{piperno2008search} can be adopted.
We will investigate these possibilities, and
benchmark and analyze the computational scaling
of the present algorithm in future.

\begin{algorithm}[H]
\caption{Partition backtrack search for canonically labeling a graph and
finding generators of its automorphism group}\label{alg}
{\bf Input:} Tensor product expression (or its corresponding graph $G$) and
its permutation symmetry group $\grp$ \\
{\bf Output:} Canonical labeling and generators of $\Aut(G)$\scriptsize
\begin{algorithmic}[1]

\Function{ToCanonicalForm}{tensor product expression, group $\grp$}
    \State Transform tensor product expression to a graph $G$,
    \State Initialize $\Pi_0$, the certificate $\Eset_0$, generating set $K=\{\;\}$
    \State Backtrack($G$, $\Pi_0$, $\grp$)
    \State Back transform $(G,\Pi_c)$ to expression
\EndFunction

\Function{Backtrack}{$G$, $\Pi$, $\grp$}

\If {$\Eset(G,\Pi)>\Eset(G,\Pi_c)$ or $(G,\Pi)$ is automorphic to a
visited node} \State \Return \EndIf

\If {$|V_i|=1\;(i=1,2,\cdots,n)$}
    \If{$\Eset(G,\Pi)=\Eset(G,\Pi_c)$}
        \State Find an automorphism $\Pi^{-1}\Pi_c$ and append it to $K$
        \State Go back to the deepest common ancestor to omit the entire
        subtree containing $\Pi$
    \Else
        \State Find a smaller $\Eset(G,\Pi)$ and update $\Pi_c=\Pi$ and $\Eset(G,\Pi_c)=\Eset(G,\Pi)$
    \EndIf
\Else
    \State Select the first non-singleton cell $v$
    \State Compute $\Stab_\grp(1)$ and the left coset representatives $\lcr$
    \For {$g\in \lcr$}
        \State Compute $\Pi^g$, split it to $\Pi^g\downarrow v_k$, and refine
$R(G,\Pi^g\downarrow v_k)$
    \EndFor
    \State Sort the refinements $\{R(G,\Pi^g\downarrow v_k)\}$ by the corresponding $\Eset(G,R(G,\Pi^g\downarrow v_k))$
    \For {$\Pi_i\in\{R(G,\Pi^g\downarrow v_k)\}$}
        \State Backtrack($G$, $\Pi_i$, $\Stab_\grp(1)$)
    \EndFor
\EndIf

\EndFunction
\end{algorithmic}
\end{algorithm}

\section{Illustrative examples}\label{results}
In this section, we start with some simple tensor products to show
how the above algorithm works in details, and then provide results for more
complicated expressions.

\subsection{Example: $I_{ij}^{ab}=t_{i}^{a}t_{j}^{b}$}
This example gives $S=(\bm{t},\bm{t})$, $\pi=\{a,i,b,j\}$,
$\grp(S)=\mathcal{S}_2(\bm{t})=\langle (13)(24)\rangle$. The
corresponding graph is constructed via $\Vset=\{1,2,3,4,5\}$ and
\begin{eqnarray}
\Eset^E(\pi)&=&
\{ \{(1,2),vir[a]\}, \{(1,3), occ[i]\},\nonumber\\
&&\;\;\{(1, 4),vir[b]\},\{(1,5), occ[j]\} \},
\end{eqnarray}
where an auxiliary external vertex (labeled by 1) has been added.
To compute the permutation symmetry
group, we can use the externally unlabeled edge set
\begin{eqnarray}
\Eset(\pi)&=&
\{\{(1,2),vir[a]\}, \{(1,3), occ[i]\},\nonumber\\
&&\;\;\{(1, 4), vir[a]\},\{(1,5), occ[i]\}\}.\label{EPi1}
\end{eqnarray}
The initial partition is
$\Pi_0=\{\{1\},\{2,3,4,5\}\}$ and accordingly the group $\grp(S)$ is
replaced by $\grp(S)=\langle (24)(35)\rangle$ via relabeling. The above
algorithm gives the following visiting sequence,
\begin{eqnarray}
\Pi_0&=&\{\{1\},\{2,3,4,5\}\}\nonumber\\
\Pi_1&=&\{\{1\},\{2\},\{3\},\{4\},\{5\}\}\nonumber\\
\Pi_2&=&\{\{1\},\{4\},\{5\},\{2\},\{3\}\}.
\end{eqnarray}
The stabilizer of the first element 2 in the non-singleton cell in
$\Pi_0$ is the trivial group $\langle e\rangle$, and the left coset
representatives are just $\{e,(24)(35)\}$. The partition $\Pi_1$ is
obtained from $\Pi_0=\{\{1\},\{2,3,4,5\}\}$ by first individualizing
$2$, $\Pi_0\downarrow 2=\{\{1\},\{2\},\{3,4,5\}\}$, and then
immediately $\mathcal{R}(G,\Pi_0\downarrow 2)=\Pi_1$ since the
subgroup of $\grp(S)$ that stabilizes $\Pi_0\downarrow2$ is the
trivial group. Then the backtrack search proceeds to the image of
the next left coset representative $\Pi_0\downarrow
4=\{\{1\},\{4\},\{5,2,3\}\}$, which again immediately leads to
$\mathcal{R}(\Pi_0\downarrow 4)=\Pi_2$ by refinement. Since
$\Eset(\pi,\Pi_1)=\Eset(\pi,\Pi_2)=\Eset(\pi)$ in Eq. \eqref{EPi1}
for the two discrete partitions $\Pi_1$ and $\Pi_2$, we have the automorphism group
$\Aut_{\grp(S)}(G(\pi))=\langle(24)(35)\rangle$, which essentially
shows $I_{ij}^{ab}=I_{ji}^{ba}$ through the homomorphism $\phi$
\eqref{Homo}, see also Theorem \ref{connectionFinal},
because for this example $\Aut_{\grp(S)}(G^E(\pi))=\langle e\rangle$ is trivial.

\subsection{Example: $t^{cba}_{kji}$}
This example illustrates how the partition backtrack works for the canonicalization
of a single tensor with only external indices. The coloring is
$\pi=\{c,b,a,k,j,i\}$. The initial partition is
$\Pi_0=\{\{1\},\{2,3,4,5,6,7\}\}$ and
$\grp(S)=\grp(\bm{t}_3)=\langle (23), (234), (56), (567) \rangle$
for the triple excitations, which is of order 36. After the following four steps
\begin{eqnarray}
\Pi_1&=&\{\{1\},\{4\},\{2,3,5,6,7\}\} \nonumber\\
\Pi_2&=&\{\{1\},\{4\},\{3\},\{2\},\{5,6,7\}\} \nonumber\\
\Pi_3&=&\{\{1\},\{4\},\{3\},\{2\},\{7\},\{5,6\}\} \nonumber\\
\Pi_4&=&\{\{1\},\{4\},\{3\},\{2\},\{7\},\{6\},\{5\}\} \nonumber
\end{eqnarray}
we arrive at $\pi_{\canon}=\{a,b,c,i,j,k\}$ and the corresponding
canonical form of tensor $t^{abc}_{ijk}$. This shows in the case
where there is no internal index, the partition backtrack algorithm
behaves as selection sorts for the indices in $\{c,b,a\}$ and $\{k,j,i\}$, respectively.

\subsection{Example: $\langle ij\|ab\rangle t_{i}^{a}t_{j}^{b}$}
This simple tensor product without any external index appears in the
coupled-cluster energy expression in terms of
antisymmetrized integrals $\bar{g}_{ij,ab}\triangleq \langle ij\|ab\rangle$.
In our convention, we have $S=(\bm{\bar{g}},\bm{t}_1,\bm{t}_1)$ and $\pi=\{i,j,a,b,a,i,b,j\}$.
The group of the antisymmetrized integral is
$\grp(\bm{\bar{g}})=\langle(12),(34),(13)(24)\rangle$ and accordingly
$\grp(S)=\grp(\bm{\bar{g}})\times\mathcal{S}_2(\bm{t}_1)= \langle
(12),(34),(13)(24),(57)(68)\rangle$, which is of order 16. The
initial partition is $\Pi_0=\{\{1,2,3,4\},\{5,6,7,8\}\}$. The
breadth-first search produces the following ordering for the
children in increasing order of $\Eset(\pi,\Pi_i)$,
\begin{eqnarray}
\Pi_1&=&\mathcal{R}(\Pi_0\downarrow 3)=\mathcal{R}(\{\{3\},\{4,1,2\},\{5,6,7,8\}\})\}\nonumber\\
&=&\{\{3\},\{4\},\{1,2\},\{5\},\{6\},\{7\},\{8\}\}\nonumber\\
\Pi_2&=&\mathcal{R}(\Pi_0\downarrow 4)=\mathcal{R}(\{\{4\},\{3,1,2\},\{5,6,7,8\}\})\}\nonumber\\&=&
\{\{4\},\{3\},\{2,1\},\{7\},\{8\},\{5\},\{6\}\}\nonumber\\
\Pi_3&=&\mathcal{R}(\Pi_0\downarrow 1)=\mathcal{R}(\{\{1\},\{2,3,4\},\{5,6,7,8\}\})\}\nonumber\\&=&
\{\{1\},\{2\},\{3,4\},\{5\},\{6\},\{7\},\{8\}\}\nonumber\\
\Pi_4&=&\mathcal{R}(\Pi_0\downarrow
2)=\mathcal{R}(\{\{2\},\{1,3,4\},\{5,6,7,8\}\})\}\nonumber\\&=&
\{\{2\},\{1\},\{4,3\},\{7\},\{8\},\{5\},\{6\}\}.\nonumber
\end{eqnarray}
Visiting $\Pi_1$ and $\Pi_2$ eventually leads to an automorphism
$(12)(34)(57)(68)$. The visit of $\Pi_3$ reveals that it produces an edge list larger than the minimal edge list found so far, thus both $\Pi_3$ and the partition
after it, viz., $\Pi_4$, can be pruned, since
$\Eset(\pi,\Pi_4)\ge\Eset(\pi,\Pi_3)>\Eset(\pi,\Pi_1)$ due to the
reordering after the initial breadth-first scan. Finally, based on
$\Eset(\pi,\Pi_1)=\{\{(1,5),vir[a]\},\{(2,7),vir[a]\},\{(3,6),occ[i]\},\{(4,8),occ[i]\}\}$,
we can relabel the tensor product as
$\bar{g}_{a_1a_2,i_1i_2}t^{a_1}_{i_1}t^{a_2}_{i_2}$, which is the target
canonical form for the input $\bar{g}_{ij,ab}t_{i}^{a}t_{j}^{b}$.
It deserves to be mentioned that in the diagrammatic
technique for coupled-cluster theory, the inverse of the order of
the automorphism group is just the weight factor associated with the
diagram\cite{paldus1975}, which will be added to the expression when
taking summations over internal indices. In the diagrammatic coupled
cluster theory, the two $\bm{t}_1$ vertices in this example are
referred as equivalent vertices, and will contribute to a factor
$1/2$. Here, our algorithm computes
the correct automorphism group $\Aut_{\grp(S)}(G(\pi))=
\langle(12)(34)(57)(68)\rangle$ with $|\Aut_{\grp(S)}(G(\pi))|=2$.

\subsection{Example: $\langle ab\|ij\rangle t_{ij}^{ab}$}
\begin{figure}[t]\centering
\resizebox{0.45\textwidth}{!}{\includegraphics{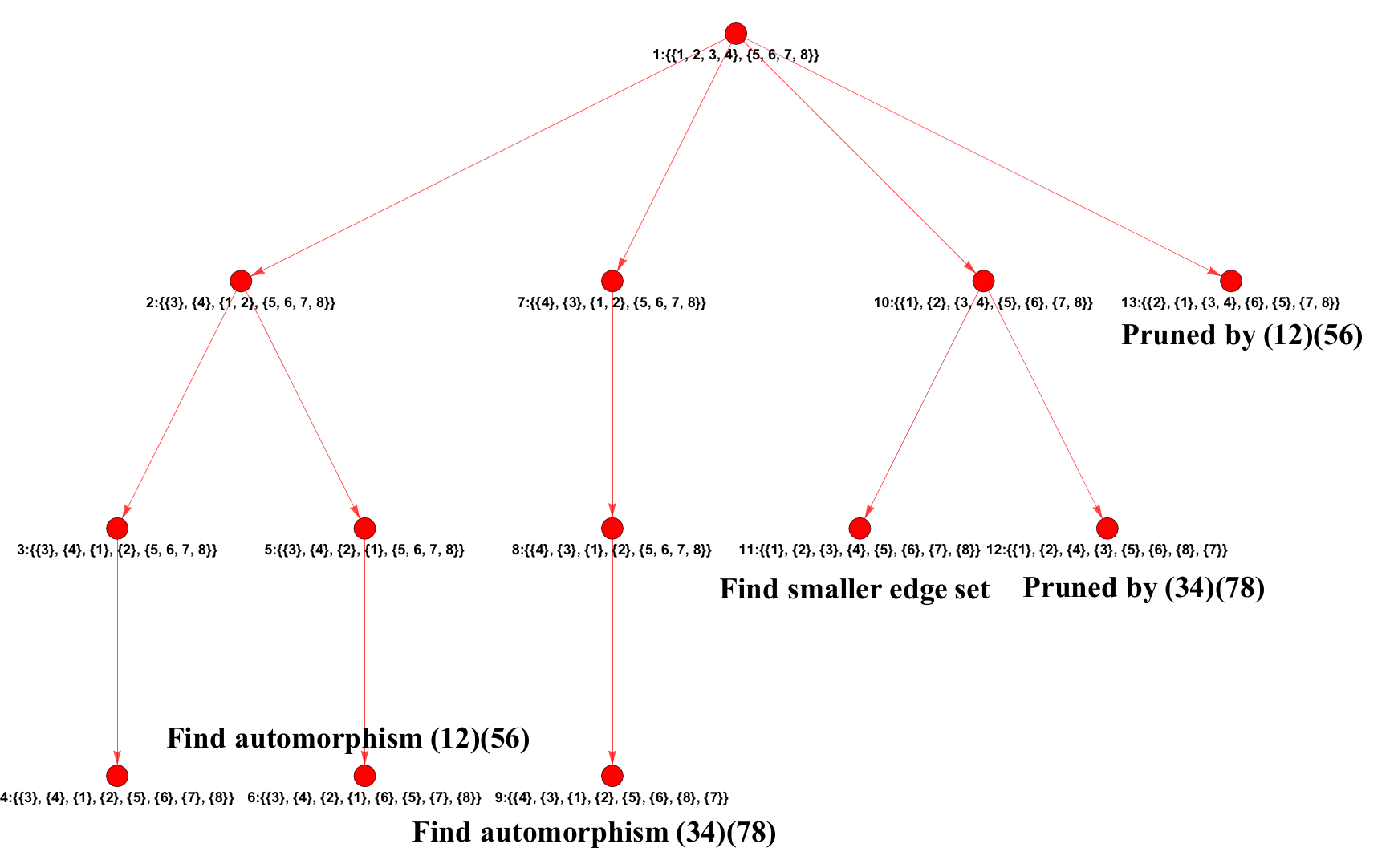}}   \\
\caption{Search tree for canonicalization of $\langle ab\|ij\rangle
t^{ab}_{ij}$.}\label{searchtreeGT2}
\end{figure}

This tensor product also appears in the coupled-cluster energy
expression. Similar to the second example, we have
$S=(\bm{\bar{g}},\bm{t}_2)$ and $\pi=\{a,b,i,j,a,b,i,j\}$. The permutation
symmetry group is
$\grp(S)=\grp(\bm{\bar{g}})\times\grp(\bm{t}_2)=\langle(12),(34),(13)(24),(56),(78)\rangle$,
which is of order 32. This example is very typical concerning with the
pruning based on automorphism. The search tree for canonicalization of
the expression is shown in Figure \ref{searchtreeGT2}.
It is seen that once the automorphism
$(34)(78)$ is found, the entire subtree at $\{\{4\}, \{3\}, \{1,
2\}, \{5, 6, 7, 8\}\}$ is pruned, since it can be mapped to the
previously visited subtree at $\{\{3\}, \{4\}, \{1, 2\}, \{5, 6, 7, 8\}\}$ by
$(34)(78)$. Besides, the entire subtree at $\{\{2\}, \{1\}, \{3,
4\}, \{6\}, \{5\}, \{7, 8\}\}$ is also completely pruned, since the
automorphism $(12)(56)$ will map it to the visited subtree at
$\{\{1\}, \{2\}, \{3, 4\}, \{5\}, \{6\}, \{7, 8\}\}$. In sum,
the automorphism group is found as $\langle (12)(56),(34)(78)\rangle$,
which is of order 4. This agree with the diagrammatic rules in the coupled-cluster theory\cite{paldus1975}, where there are two pairs of equivalent internal lines,
which contributes to a weight factor $1/2^2=1/4$.

\subsection{Example: $A_{i_{\sigma(1)}i_{\sigma(2)}\cdots
i_{\sigma(n)}}A_{i_{\tau(1)}i_{\tau(2)}\cdots i_{\tau(n)}}$ with
$\sigma,\tau\in \mathcal{S}_n$ and $\grp(\bm{A})=\mathcal{S}_n$}
The final example is used to illustrate the performance of
the partition backtrack algorithm for challenging cases, viz., tensor products
with large automorphism groups, implying that there are many partitions
with the same edge sets. Such example was identified
as the worst case for the traditional Bulter-Portugal algorithm,
which leads to an exponential cost\cite{niehoff2018faster}.

\begin{figure}\centering
\resizebox{0.5\textwidth}{!}{\includegraphics{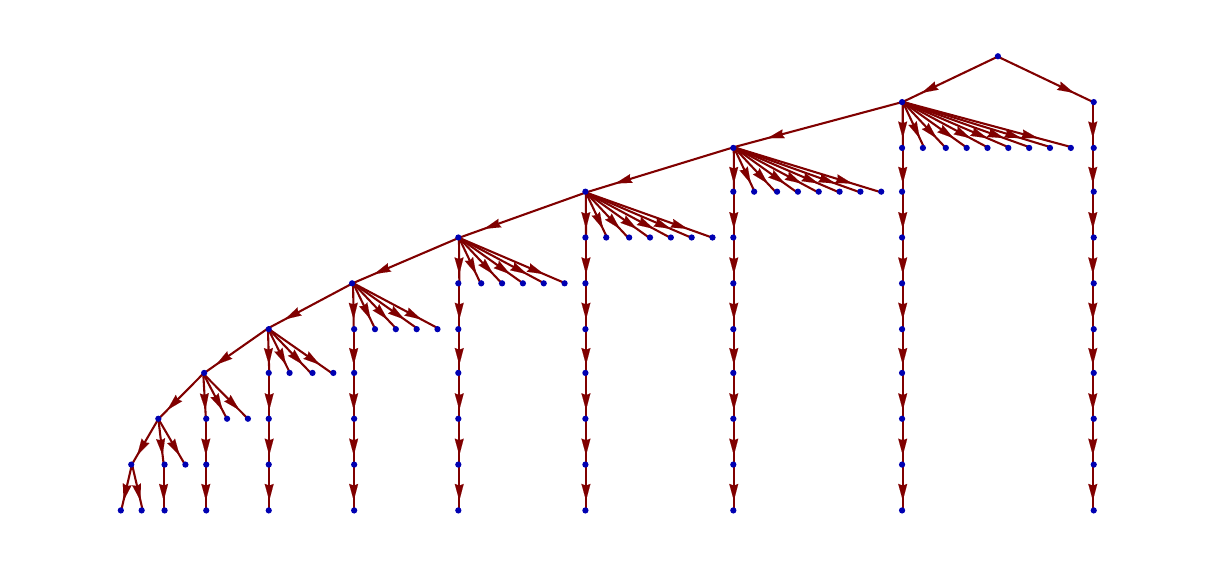}}\\
\caption{Search tree for $A_{i_{1}i_{2}\cdots
i_{10}}A_{i_{1}i_{2}\cdots i_{10}}$ with
$\grp(\bm{A})=\mathcal{S}_{10}$}\label{searchtreeS5}
\end{figure}

The search tree for $n=10$ is displayed
in Figure \ref{searchtreeS5}. In general, the size of search tree for such product
in our algorithm can be found as $n^2+2$. This is in sharp comparison with
the enormous size of the automorphism group in this case,
which is in general,
\begin{eqnarray}
\Aut_{\grp(S)}(G)
&=&\langle (1,2)(n+1,n+2),\nonumber\\
&&\;(2,3)(n+2,n+3),\cdots,\nonumber\\
&&\;(n-1,n)(2n-1,2n),\nonumber\\
&&\;(1,n+1)(2,n+2)\cdots (n,2n)\rangle,
\end{eqnarray}
of order $2(n!)$, and $|\Aut_{\grp(S)}(G)
|=7257600$ for $n=10$.
As shown in Figure \ref{searchtreeS5}, only the
leaves corresponding to the generators of the
automorphism group are examined, while
all other branches are pruned by automorphisms.
This pruning lead to a polynomial scaling $O(n^2)$ in the size
of search tree with respect to $n$.

In sharp contrast, Ref. \cite{niehoff2018faster} shows that the existing algorithm
will take a full day for $n=12$ (see Sec 3.4 in Ref. \cite{niehoff2018faster}
for worst-case complexity analysis).
[NB: The same applies to the built-in canonicalization function  \texttt{TensorReduce} in \textit{Mathematica}.] The present algorithm
took 3 seconds for $n=12$ due to its polynomial scaling in this case,
even with a very preliminary implementation.
This reveals that the present graph based reformulation of the canonicalization problem
provides a very promising framework for future improvements.
We will investigate the possible improvements mentioned in Sec.
\ref{sec:improvement} and provide an optimized implementation in future.

\section{Conclusion and outlook}\label{conclusion}
The present work provides a complete classification of tensor
product expressions by means of equivalence relation and group chain. We
provide a rigorous definition of canonical form for the tensor products based on the
classification theory, a graphical presentation for tensor products, a very promising
partition backtrack algorithm to compute the canonical form and automorphism group,
and an explicit construction of permutation symmetry group of the resulted
tensor $Z^E$ \eqref{tensorProduct}. These solve the four fundamental questions
Q1-Q4 raised in the introduction. We note that
the automorphism group $\Ker\phi$ and the permutation symmetry group $\grp(\Omega_E(\pi))$ contain very interesting information about the tensor products. In particular,
when augmented with more general definition of tensor symmetry (with signs or general phases), more information can be extracted from these groups. A particular important
case is that when considering tensors with certain antisymmetry,
if there is a minus sign factor associated with an element
in $\Ker\phi$ found by the algorithm,
then the resulted tensor products can be concluded as zero,
without any numerical calculations. A simple
example is the product $A_{i_1i_2}B_{i_1i_2}$, where $A$ is
antisymmetric and $B$ is symmetric, resulting a vanishing contraction
by symmetry. Further applications of the present classification theory and graphical canonicalization algorithms in automatic derivation and simplification of general tensor product expressions, in particular, the exploration of the use of the automorphism group $\Ker\phi$ and permutation symmetry group $\grp(\Omega_E(\pi))$, will be presented in future.


%

\end{document}

%% file: conventions.tex
\usepackage{amsthm}
\newtheorem{dfn}{Definition}
\newtheorem{lem}{Lemma}
\newtheorem{thm}{Theorem}

\newcommand{\grp}{\mathcal{G}}
\newcommand{\subgrp}{\mathcal{H}}

\newcommand{\lcr}{\mathcal{L}}

\newcommand{\Aut}{\mathrm{Aut}}
\newcommand{\Ker}{\mathrm{Ker}}

\newcommand{\Stab}{\mathrm{Stab}}
\newcommand{\Vset}{\mathcal{V}}
\newcommand{\Eset}{\mathcal{E}}
\newcommand{\Class}[1]{\llbracket #1 \rrbracket}
\newcommand{\canon}{\mathrm{canon}}